\documentclass[prd,english,preprintnumbers,amsmath,amssymb,nofootinbib,superscriptaddress,twocolumn]{revtex4-1}

\pdfoutput=1

\usepackage{mathtools}
\usepackage{amsfonts}
\usepackage{mathrsfs}
\usepackage{bbm}
\usepackage{slashed}

\usepackage{graphicx}
\usepackage[dvipsnames]{xcolor}
\usepackage{array}

\usepackage{xspace}
\usepackage{siunitx}
\usepackage{hyperref}

\usepackage{xifthen}
\usepackage{dsfont}
\usepackage{appendix}
\usepackage{enumitem}
\usepackage{booktabs}
\usepackage{units}

\usepackage[latin1]{inputenc}
\usepackage{array}

\graphicspath{
	{./}
}

\def\0#1#2{\frac{#1}{#2}}

\def\s0#1#2{\mbox{\small{$ \frac{#1}{#2} $}}}

\def\CC{{\mathcal C}}

\newcommand{\I}{\mathrm{i}}
\newcommand{\be}{\begin{eqnarray}}
\newcommand{\ee}{\end{eqnarray}}

\newcommand{\nn}{\nonumber }
\newcommand{\fslash}{\hspace*{-0.2cm}\slash }

\newcommand{\beq}{\begin{equation}}
\newcommand{\eeq}{\end{equation}}
\newcommand{\bea}{\begin{eqnarray}}
\newcommand{\eea}{\end{eqnarray}}

\newcommand{\LQCD}{\Lambda_{\text{QCD}}}
\newcommand{\ksb}{k_{\text{SB}}}

\newcommand{\psib}{\bar{\psi}}

\newcommand{\delfourn}[1]{\;(2\pi)^4\delta^{(4)}\left(#1\right)}

\def\0#1#2{\frac{#1}{#2}}

\def\eq#1{\eqref{#1}}

\hypersetup{
	colorlinks,
	linkcolor={red!75!black},
	citecolor={blue!75!black},
	urlcolor={blue!75!black},
	pdftitle={From quarks and gluons to color superconductivity at supranuclear densities},
	pdfauthor={Braun, Schallmo},
	pdfkeywords={Renormalization Group} {Color superconductivity}
	bookmarksopen=true,
	bookmarksopenlevel=2,
	bookmarksnumbered=true
}

\usepackage{babel}
\makeatother
\begin{document}

\title{From quarks and gluons to color superconductivity at supranuclear densities}

\author{Jens Braun}
\affiliation{Institut f\"ur Kernphysik, Technische Universit\"at Darmstadt, 
D-64289 Darmstadt, Germany}
\affiliation{ExtreMe Matter Institute EMMI, GSI, Planckstra{\ss}e 1, D-64291 Darmstadt, Germany}
\author{Benedikt Schallmo} 
\affiliation{Institut f\"ur Kernphysik, Technische Universit\"at Darmstadt, 
D-64289 Darmstadt, Germany}

\begin{abstract}
We study the emergence of color superconductivity in the theory of the strong interaction at supranuclear densities. 
To this end, we follow the renormalization group (RG) flow of dense strong-interaction 
matter with two massless quark flavors from the fundamental quark and gluon degrees of freedom 
at high energies down to the non-perturbative low-energy regime which is found to be governed by the dynamical formation of diquark states. 
With the strong coupling at the initial RG scale as the only input parameter, we compute the (chirally symmetric) scalar diquark condensate 
and analyze its scaling behavior over a wide range of the quark chemical potential. 
Approximations entering our computations are critically assessed. 
Since our approach naturally allows us to study the scale dependence of couplings, we also monitor the strength of couplings appearing in 
low-energy models of dense strong-interaction matter. The observed dependence of these couplings on the quark 
chemical potential may help to amend model studies in the future. Finally, we estimate the speed of sound of 
dense QCD matter. Our results indicate that the speed of sound exceeds the value 
of the noninteracting quark gas at high densities and even increases as the density is decreased, across a wide range, 
suggesting the existence of a maximum at supranuclear densities.
\end{abstract}
\maketitle

%
\section{Introduction}
There is interest in the properties of Quantum Chromodynamics (QCD) at supranuclear densities  
ever since the first discussion of the possible existence of color-superconducting ground states 
in the 1970s, see Ref.~\cite{Bailin:1983bm} for an early review. However, the properties of such states remained elusive for a long time. 
In the late 1990s, it was then found that the formation of sizeable pairing gaps in 
color-superconducting phases may considerably affect the dynamics of QCD at low temperatures, 
see Refs.~\cite{Rajagopal:2000wf,Alford:2001dt,Buballa:2003qv,Shovkovy:2004me,Alford:2007xm,Fukushima:2010bq,Fukushima:2011jc,Anglani:2013gfu,Schmitt:2014eka,Baym:2017whm} for reviews. 

More recently, the interest in the properties of dense strong-interaction matter received a significant boost because of the first detection of the 
gravitational-wave signal of a neutron-star merger~\cite{TheLIGOScientific:2017qsa,Abbott:2018wiz}, ongoing
missions aiming at first direct neutron-star radius measurements~\cite{Watts:2016uzu,NICER,NICER2,miller2021radius,riley2021nicer,Raaijmakers:2021uju}, as well as 
precise mass measurements of heavy neutron stars~\cite{Demorest10,Antoniadis13,Fonseca2016,2019arXiv190406759C}.
These breakthroughs provide important constraints for the equation of state (EOS) of strong-interaction matter, see 
Ref.~\cite{Huth:2020ozf} for a recent analysis. 
Quantitative theoretical results for the EOS of dense strong-interaction matter are therefore indeed urgently needed in view of this 
tremendous progress made in the observation of 
neutron stars. In addition, constraints on the EOS can be obtained from heavy-ion collisions~\cite{Danielewicz:2002pu}.
Nevertheless, a reliable description of the properties and dynamics of strong-interaction matter over a wide range of densities and temperatures still represents 
a formidable challenge, from an observational, experimental, and theoretical standpoint. 

Presently, studies based on chiral effective field theory (EFT) interactions (see, e.g., Ref.~\cite{Epelbaum:2008ga} for a review) 
set benchmarks and yield strong constraints for the EOS in the low-density regime~\cite{Hebeler:2013nza,Leonhardt:2019fua}, see Ref.~\cite{Hebeler:2020ocj} for a recent review.  
For low to moderate densities, functional renormalization group (fRG) studies of 
nucleon-meson~\cite{Berges:1998ha,Drews:2014wba,Drews:2014spa}
and quark-meson models~\cite{Tripolt:2017zgc,Otto:2019zjy,Otto:2020hoz} aiming 
at the EOS at low temperatures are also available.
At very high density, constraints for the EOS come from perturbative QCD (pQCD) studies~\cite{Freedman:1976xs,Freedman:1976ub,Baluni:1977ms,Kurkela:2009gj,Fraga:2013qra,Fraga:2016yxs,Gorda:2018gpy}. 
However, in the broad intermediate density regime, where both the chiral and the pQCD expansion are expected to break 
down, much less is known about the dynamical degrees of freedom and their interactions, resulting in large 
uncertainties for the EOS and other quantities, such as the speed of sound of dense matter.
In this density regime, which is still relevant for astrophysical applications, QCD is widely expected to be governed by a 
color-superconducting ground state (see Refs.~\cite{Rajagopal:2000wf,Alford:2001dt,Buballa:2003qv,Shovkovy:2004me,Alford:2007xm,Fukushima:2010bq,%
Fukushima:2011jc,Anglani:2013gfu,Schmitt:2014eka,Baym:2017whm} for reviews). 

Renormalization Group (RG) approaches have played and are still playing an outstanding role in the analysis of the symmetry-breaking patterns and the emergence of 
color-superconducting ground states in dense matter~\cite{Son:1998uk,Schafer:1999jg,Hsu:1999mp,Braun:2018bik,Braun:2019aow}. Indeed, 
since systems of this kind represent a multi-scale problem, RG approaches are very well suited. 
Recently, an analysis of the RG flow of
gluon-induced four-quark interaction channels in a Fierz-complete setting for two massless quark flavors has been performed 
to gain a deeper insight into symmetry breaking patterns underlying QCD 
over a wide range of densities at low and intermediate temperatures~\cite{Braun:2019aow}. 
There, it was found that the scalar-pseudoscalar 
interaction channel dominates the dynamics for small chemical potentials. 
Increasing the chemical potential, 
a (small) range of chemical potentials opens up with many interaction channels of roughly equal strength, indicating that the structure of the 
ground state may be very complicated in this regime.
Increasing the chemical potential further, it was then observed that the diquark channel becomes most dominant, suggesting the formation of a 
chirally symmetric diquark condensate associated with pairing of the two-flavor color-superconductor (2SC) 
type. This observation is in accordance with early studies~\cite{Alford:1997zt,Rapp:1997zu,Schafer:1998na,Berges:1998rc}, including
first-principles calculations which exploit the fact that the coupling effectively becomes small in the high-density limit owing to 
asymptotic freedom~\cite{Son:1998uk,Schafer:1999jg,Pisarski:1999bf,Pisarski:1999tv,Brown:1999aq,Evans:1999at,Hong:1999fh}. 

The RG analysis of the symmetry-breaking patterns in Ref.~\cite{Braun:2019aow} laid the ground for a subsequent computation of constraints 
from quark-gluon dynamics for the EOS of isospin-symmetric two-flavor QCD over a wide range of densities~\cite{Leonhardt:2019fua}.
Remarkably, towards the nucleonic low-density regime, the results from this EOS study are impressively consistent with those from calculations based on chiral EFT 
interactions. Moreover, the RG study of the EOS in Ref.~\cite{Leonhardt:2019fua} predicts the 
emergence of a maximum in the speed of sound at supranuclear densities which appears to be tightly 
connected to the formation of a diquark gap. Interestingly, this maximum exceeds the asymptotic high-density value of the speed of sound. 
However, its exact position in terms of the density has not yet been determined conclusively. With respect to astrophysical 
applications, it is worth noting that the analysis of constraints 
from neutron-star masses also strongly suggests the existence of a maximum of the speed of sound for neutron-rich matter~\cite{Bedaque:2014sqa,Tews:2018kmu,Greif:2018njt,Annala:2019puf,Huth:2020ozf}.  

With our present work, we aim at laying the field-theoretical foundation for new first-principles 
studies of the EOS of dense QCD matter. As a first application, we shall demonstrate that -- starting from the fundamental quark and gluon degrees of freedom 
at high energies -- our RG approach allows us to 
study the dynamical formation of diquarks in the low-energy limit. 
In a next step, for example, this can be used 
to narrow down the (systematic) uncertainties of the thermodynamic quantities computed in Ref.~\cite{Leonhardt:2019fua}, in particular 
those of the EOS and the position of the maximum of the speed of sound. 
Still, the analysis of the RG flows presented in this work already allows us to gain 
an insight into the dynamics of dense QCD matter over a wide range of chemical potentials, as we shall show by computing the diquark gap.

The present work is organized as follows: In Sec.~\ref{sec:formalism}, we discuss the formalism underlying our RG analysis 
of dense QCD matter. This includes a discussion of possible extensions required for computations of the EOS and also makes connections 
to our previous study of the EOS of dense matter~\cite{Leonhardt:2019fua}. 
The RG flow of dense QCD matter is then analyzed in detail in Sec.~\ref{sec:RGhighdens}. There, 
we also present our results for the (chirally symmetric) scalar diquark condensate as a function of the quark chemical potential. 
In Sec.~\ref{sec:LEM}, we finally discuss 
implications of our RG study for low-energy models of dense strong-interaction matter and for the 
speed of sound as a specific example for a phenomenologically important thermodynamic quantity. 
Our conclusions and a brief outlook can be found in Sec.~\ref{sec:conc}.

\section{Formalism}
\label{sec:formalism}
\subsection{Effective Action}
For our analysis of the properties of QCD at intermediate and also high densities, we employ the Wetterich equation~\cite{Wetterich:1992yh} 
which is an RG equation for the quantum effective action~$\Gamma$. Within this framework, the effective action depends on a so-called 
RG ``time" $t=\ln(k/\Lambda)$ where~$k$ is the RG scale and~$\Lambda$ may be chosen to be the scale at which the 
initial condition~$\Gamma_{k=\Lambda}$ for the scale-dependent effective action~$\Gamma_k$ is fixed. In our present study, 
the initial condition is given by the classical (Euclidean) QCD action~$S$ for two massless quark flavors coming in three colors: 
\be
S = \int {\rm d}^4x\left\{ 
\frac{1}{4}F_{\mu\nu}^{a}F_{\mu\nu}^{a}
+ \bar{\psi}\left(
{\rm i}\slashed{\partial} + \bar{g}\slashed{A} - {\rm i}\gamma_0\mu \right)\psi
\right\}\,.
\label{eq:qcd}
\ee
Here,~$\bar{g}$ is the bare gauge coupling and~$\mu$ is the quark chemical
potential. For the values of~$\mu$ considered in this work, 
we choose~$\Lambda\gg \mu$ to ensure that the RG flow is initialized in the perturbative high-energy regime. 
The gluon fields~$A_{\mu}^a$ come with Lorentz (greek letters) and color (roman letters) indices and enter the definition of the
field-strength tensor~$F_{\mu\nu}^a =\partial_\mu A^a_\nu-\partial_\nu A^a_\mu + \bar{g} f^{abc} A_\mu^b A_\nu^c$ ($a=1,...,8$).
Moreover, they are coupled to the quark fields~$\psi$ via 
the quark-gluon vertex, see Eq.~\eqref{eq:qcd}. 
Note that the quark fields~$\psi$ carry color and flavor components. 

The quark-gluon vertex generates a plethora of interaction channels. With respect to studies 
of ground-state properties, quark self-interaction channels are of particular importance as they can be directly 
related to the order-parameter potential of QCD. More specifically, the quark-gluon vertex 
induces four-quark interactions already at the one-loop level via two-gluon exchange. Schematically, this leads to 
corrections of the effective action of the following form:
\be
\Delta\Gamma = \int {\rm d}^4x\,\sum_i\bar{\lambda}_{i}(\bar{\psi}{\mathcal O}_i\psi)^2\,.
\label{eq:deltag}
\ee
Here,~${\mathcal O}_i$ determines the color, flavor, and Dirac
structure of the four-quark vertex. Note that, in contrast to low-energy model studies, 
the four-quark couplings~$\bar{\lambda}_i$ are not free parameters but generated from 
fundamental quark-gluon interactions,~$\bar{\lambda}_i\sim \bar{g}^4$. Higher quark self-interactions 
are parametrically suppressed at high momentum scales. For example, eight quark-interactions scale as~$\sim\bar{g}^8$.  
However, following the RG flow from high to low momentum scales, such higher-order interaction channels then 
become increasingly important. In fact, in regimes where the symmetry is broken spontaneously, 
eight quark interactions determine the masses of bound states of two quarks. We shall come back to this below. 
In particular, we shall discuss the relevance of eight-quark interactions at different scales in Sec.~\ref{sec:LEM}, which may 
also provide useful information for the construction of low-energy models at intermediate and high densities. 

Still, already an analysis of the RG flow of gluon-induced four-quark interactions in the pointlike limit (``zero-momentum projection") can
provide us with an important insight into the symmetry-breaking patterns 
over a wide range of temperatures and quark chemical potentials, see Ref.~\cite{Braun:2011pp} for an introduction. 
In fact, this has been successfully demonstrated for 
QCD in the vacuum limit~\cite{Gies:2005as}, at finite temperature~\cite{Braun:2005uj,Braun:2006jd}, and over a wide range 
of chemical potentials~\cite{Braun:2019aow}. In the latter study, it has been found within a Fierz-complete two-flavor setting 
that the scalar-pseudoscalar channel is most dominant at low densities, in accordance with full QCD RG-flows in the 
vacuum limit~\cite{Mitter:2014wpa,Cyrol:2017ewj}. At large chemical potentials, which are at the heart of the present work,
the diquark channel~$\sim (\psib_b\tau_2\epsilon_{abc}\gamma_5\CC \psib^T_c)(\psi^T_d\CC\gamma_5\tau_2\epsilon_{ade}\psi_e)$ 
is then dynamically rendered the most 
dominant channel, suggesting the formation of a chirally symmetric diquark
condensate associated with pairing of the two-flavor color-superconductor (2SC) 
type~\cite{Braun:2019aow}.\footnote{Here, $\tau_2$ is the second Pauli matrix and, in color space, it is summed over the totally 
antisymmetric tensor~$\epsilon_{abc}$. 
Moreover, we have introduced $\mathcal{C}=\I \gamma_2 \gamma_0$.}
This is in accordance with early studies of dense QCD~\cite{Rapp:1997zu,Alford:1997zt,Berges:1998rc,Pisarski:1999tv,Pisarski:1999bf,Schafer:1999jg}. 

Although studies of the RG flow of four-quark interactions in the pointlike approximation provide a deep insight into symmetry-breaking patterns 
and their dependence on external control parameters, they are restricted to scales~$k \geq k_{\text{SB}}$, where the scale~$k_{\text{SB}}$ 
is associated with spontaneous symmetry breaking, such as chiral symmetry breaking or~${\rm U}(1)_{\text{V}}$ symmetry breaking. 
In such a setting, symmetry breaking is indicated by a specific four-quark channel approaching criticality 
associated with a divergence of the corresponding coupling 
at the scale~$k_{\text{SB}}$. 
Below this scale, the dynamics is governed by the formation of condensates. 
However, an analysis of the ground-state properties of QCD in this low-energy regime~$k < k_{\text{SB}}$ requires to go 
beyond the pointlike limit and to resolve the momentum dependences of the quark correlation functions. Indeed, 
information on bound-state and condensate formation is encoded in the momentum structure of the quark correlation functions. 
Such momentum dependences can be conveniently resolved by employing a Hubbard-Stratonovich transformation of at least 
the most dominant four-quark interaction channel. For example, as demonstrated in Ref.~\cite{Leonhardt:2019fua}, one may 
perform such a transformation of gluon-induced four-quark interactions at a given scale~$\Lambda_0 > k_{\text{SB}}$, which then 
gives access to the low-energy regime. However, this introduces a dependence of the effective action on the scale~$\Lambda_0$ which 
is reflected in an uncertainty for the results for low-energy observables, see Ref.~\cite{Leonhardt:2019fua} for a discussion in the context of dense QCD. 
The dependence on this artificial scale~$\Lambda_0$ can be removed by employing the so-called dynamical hadronization
technique~\cite{Gies:2001nw,Gies:2002hq,Pawlowski:2005xe,Gies:2006wv,Floerchinger:2009uf,Braun:2014ata,Fu:2019hdw}, see also
Ref.~\cite{Fukushima:2021ctq} for recent developments regarding the study of quark composites.
Loosely speaking, this technique implements continuous
Hubbard-Stratonovich transformations of four-quark interactions in the RG
flow and thereby allows us to {\it continuously} follow the RG flow from the classical QCD action at high-momentum scales 
down to the deep infrared regime which is governed by the formation of bound states and condensates. We shall 
apply this technique in the following.

The present work should be viewed as the next step in a series of studies~\cite{Braun:2017srn,Braun:2018svj,Braun:2018svj,Braun:2019aow,Leonhardt:2019fua,Braun:2020bhy}. 
However, we do not aim at quantitative studies of thermodynamic quantities and low-energy observables. We rather 
aim at setting the methodological stage for subsequent new quantitative computations in this series. 

Let us now be specific and construct our ansatz for the scale-dependent effective action~$\Gamma_k$ which underlies our present study of dense QCD matter. 
As in our previous works, see, e.g., Ref.~\cite{Braun:2019aow}, we rely on the background field approach 
to gauge theories~\cite{Abbott:1980hw,Abbott:1981ke} within background covariant gauges and  
employ the background field approximation which has been worked out in detail for applications in perturbative 
as well as non-perturbative settings over many years by now,
see, e.g., Refs.~\cite{Reuter:1993kw,Reuter:1997gx,Litim:1998nf,Freire:2000bq,Litim:2002hj,Gies:2002af,%
Braun:2007bx,Braun:2010cy,Reinosa:2014ooa,Reinosa:2014zta} and, for a recent detailed fRG review on this aspect, see Ref.~\cite{Dupuis:2020fhh}. 
In this approach, the so-called background field effective action inherits gauge invariance from gauge transformations of an  
auxiliary background field. The equivalence of this invariance with the actual physical gauge invariance follows from the on-shell background independence 
of this approach and the Slavnov-Taylor identities, where the background independence is encoded in Nielsen identities. With these identities, it 
can then be shown that the correlation functions associated with the background field are indeed related to elements of the $S$-matrix~\cite{Abbott:1983zw}. In fRG studies, however,  
the regulator functions for fields carrying a net color charge break gauge invariance explicitly and, 
as a consequence, the independence of the auxiliary background field is also lost. This 
eventually leads to modifications of the Slavnov-Taylor and the Nielsen identities. Note that the latter also monitor the difference of correlation functions 
associated with the background field and those associated with the fluctuation field. In general, the construction of a manifestly gauge-invariant 
effective action in the spirit of the background-field approach may therefore be nontrivial within the fRG framework. 
In the present work, we treat the gauge sector as 
developed and discussed in detail in Refs.~\cite{Reuter:1993kw,Reuter:1997gx,Gies:2002af,Gies:2005as,Braun:2006jd}. More specifically, 
manifest gauge invariance of the solution in these studies is maintained by identifying the full gauge field with the background field 
in the RG flow. Thus, in the following, we assume that the 
background-field two-point function can be identified with the one of the fluctuation field in the flow, which is an approximation. 
For a treatment of the difference of these two quantities, we refer the reader to Ref.~\cite{Pawlowski:2001df}. This approximation 
entails that the RG flow is no longer closed~\cite{Litim:2002xm} and only some constraints imposed by the modified Slavnov-Taylor identities 
are satisfied. As in previous works~\cite{Reuter:1993kw,Reuter:1997gx,Gies:2002af,Gies:2005as,Braun:2006jd}, we shall assume 
that corrections due to this approximation are subleading, which is at least reasonable in the 
(semi-)perturbative regime above the symmetry breaking scale~$k_{\text{SB}}$. 
A detailed discussion of these issues can be found in Ref.~\cite{Dupuis:2020fhh}.
In any case, the advantage of our present approach is that it equips us with a gauge-invariant 
approximate solution of the effective action.

Since we would like to study the RG flow from the perturbative high-momentum regime down to  
the low-energy regime governed by the formation of bound states of quarks, we basically employ a combination of the classical QCD action given in Eq.~\eqref{eq:qcd} 
and an ansatz for the low-energy sector associated with complex-valued scalar diquark fields~$\Delta_a$ describing quark composites of the form~$\sim (\psi^T_b\CC \gamma_5\tau_2\epsilon_{abc}\psi_c)$:
{\allowdisplaybreaks
\begin{widetext}
\be
\Gamma_k &=& \int\text{d}^4x\,\bigg\{ \psib_b \left({\rm i}\gamma_{\mu}D_{\mu}^{bc} - {\rm i} \mu\gamma_0\right)\psi_c
+ Z_{\Delta} \left( D_{\mu}^{ca} \Delta_a\right) \left( D_{\mu}^{cb} \Delta_b\right)^{\ast} 
 + 2 \mu  Z_{\Delta} \left(\Delta_a \left(D_{0}^{ab}\Delta_b\right)^{\ast} \!-\! \Delta^{\ast}_a \left(D_{0}^{ab}\Delta_b \right)\right)
 \! -\! 4\mu^2  Z_{\Delta} \Delta^\ast_a\Delta_a
\nn\\
&&  
\qquad\qquad\qquad 
+\frac{1}{2}\bar{\lambda}_\text{csc}(\psib_b\tau_2 {\rm i}\epsilon_{abc}\gamma_5\CC \psib^T_c)(\psi^T_d\CC\gamma_5\tau_2 {\rm i}\epsilon_{ade}\psi_e)
+ \frac{1}{2}  {\rm i}\bar{h}(\psi^T_b\CC\gamma_5\tau_2\Delta_a \epsilon_{abc}\psi_c) - \frac{1}{2} {\rm i}\bar{h}(\psib_b\gamma_5\tau_2\Delta^\ast_a \epsilon_{abc}\CC \psib^T_c)
\nn\\
&&\qquad\qquad\qquad\qquad 
+ \bar{m}^2 \Delta^\ast_a\Delta_a+ \bar{\lambda}_{\Delta}(\Delta^\ast_a\Delta_a)^2
+ \frac{1}{4}Z_A F_{\mu\nu}^{a}F_{\mu\nu}^{a}  
\bigg\} + \Delta\Gamma_{\text{gf}} + \Delta\Gamma_{\text{gh}}
\,.
\label{eq:gammak}
\ee
\end{widetext}
Here},~$D_{\mu}^{bc} = {\partial_{\mu}}\delta^{bc} - \I \bar{g} A_{\mu}^{a}T^a_{bc}$ and~$a,b,c$ are color indices. 
We have suppressed flavor indices for readability. 
Note that we do not take into account the running of the 
wavefunction renormalization of the quark fields in our present exploratory study since it depends only mildly on the RG 
scale, at least at small densities~\cite{Gies:2002hq,Braun:2008pi,Braun:2014ata,Mitter:2014wpa,Rennecke:2015eba,Cyrol:2017ewj,Fu:2019hdw}.

The diquark ($\Delta^{\ast}_a$)/antidiquark  ($\Delta_a$) fields appearing in Eq.~\eqref{eq:gammak} transform as an antitriplet/triplet in color space. 
Note that we include only these fields as effective low-energy degrees of freedom. This is motivated by the fact that the diquark channel has been found to be the most dominant 
interaction channel for $\mu \gtrsim 350\,\text{MeV}$ in a Fierz-complete study of gluon-induced four-quark interaction channels~\cite{Braun:2019aow}. Other four-quark 
channels, such as the scalar-pseudoscalar interaction channel associated with pion dynamics, have been found to be clearly subdominant in this regime, provided that the 
${\rm U}(1)_{\rm A}$ symmetry is broken explicitly.  
The unspecified quantities $\Delta\Gamma_{\text{gf}}$ and $\Delta\Gamma_{\text{gh}}$ in Eq.~\eqref{eq:gammak} are the standard background-field gauge-fixing and ghost term, 
respectively. In all explicit calculations, we have restricted ourselves to Feynman gauge for convenience. 

A few comments are still in order at this point: In this work, we are aiming at a study of dense strong-interaction matter. 
To this end, we employ the diquark field as an effective degree of freedom to analyze the properties of the 
ground state. Since the diquark field is not a color-neutral object, the dynamical generation of a finite expectation 
value of this field would break the~$\text{SU}(3)$ 
color symmetry and therefore gauge invariance. Of course, it is known that local gauge invariance cannot be broken~\cite{Elitzur:1975im}. 
Moreover, the diquarks are effective degrees of freedom 
which do not even need to be asymptotic states in the spectrum. In any case, in {\mbox{(color-)}superconducting systems, 
the physics is governed by the formation of a gap in the spectrum of 
fermionic excitations at the Fermi surface and the existence of such a gap is a gauge-invariant statement. 
The description of the formation of this gap in the fermionic excitation 
spectrum in terms of a diquark condensate within a fixed gauge, 
which effectively breaks the gauge symmetry, is only a convenient choice to get access to the low-energy 
dynamics~\cite{Rajagopal:2000wf}. 
In this work, we expand the effective action in the quantity~$\Delta^\ast_a\Delta_a$ 
(summation over~$a$ is tacitly assumed), which is a gauge-invariant object. The gap in the fermionic spectrum 
is also constructed from this quantity. In practice, we employ a homogeneous background for the expansion 
and eventually evaluate the flow equations on a specific background configuration. This configuration 
is chosen to point into the $3$-direction in color space for convenience, which may possibly
lead to a residual dependence of our results for the gap on this choice. 
In future studies, our presently employed convenient approach to study the physics of dense QCD 
matter may be ``outperformed" by directly computing the full momentum 
dependence of fermonic correlation functions in a vertex expansion and searching for 
signatures of a gap in these quantities, without relying on the  
use of diquark fields as auxiliary degrees of freedom. 
However, this is beyond the scope of the present work. 
We add that, in principle, similar issues are  
encountered in the description of mass generation in the electroweak sector of the 
Standard Model~\cite{PhysRev.130.439,Englert:1964et,Higgs:1964ia,Higgs:1964pj,Guralnik:1964eu}.

Of course, by construction, our ansatz for~$\Gamma_k$ does not allow for a study of the transition from a color-superconducting phase 
at intermediate and high densities to a phase governed by spontaneous chiral symmetry breaking at low densities. Therefore, our present 
work focusses on the intermediate and high density regime. Note that a quantitative analysis of the regime associated with the aforementioned transition is anyhow 
complicated by the fact that many four-quark interaction channels have been found to be of roughly the same strength in this regime~\cite{Braun:2019aow}. 
This suggests that the ground state of QCD may exhibit a very complicated structure in this transition regime. We add that, close to the nucleonic low-density regime, the dynamics may 
even be governed by quarkyonic matter~\cite{McLerran:2018hbz}. 

\subsection{RG Flow Equations}
\label{subsec:rgfloweqs}
Let us begin our discussion of the RG flow by explaining 
the structure of our ansatz for the scale-dependent effective action~$\Gamma_k$ in more detail. The initial condition for~$\Gamma_k$ 
at the scale~$k=\Lambda \gg\mu$ 
is assumed to be given by the classical QCD action~\eqref{eq:qcd}. Therefore, the values of the four-quark coupling~$\bar{\lambda}_\text{csc}$, the quark-diquark coupling~$\bar{h}$, 
the bosonic wavefunction renormalization factor~$Z_{\Delta}$, and the four-diquark coupling~$\bar{\lambda}_{\Delta}$ 
should be set to zero at the initial RG scale~$\Lambda$.\footnote{We also refer to Subsec.~\ref{subsec:sf} for a discussion of the initial 
conditions and the RG flow at high momentum scales.}
This choice for~$Z_{\Delta}$ implies that we have $m^2=\bar{m}^2/Z_{\Delta}\to \infty$ for the renormalized mass parameter of the diquarks for~$k\to\Lambda$. Thus, 
the diquark fields are indeed not dynamical 
degrees of freedom at high-momentum scales. Their emergence in the low-energy regime of dense QCD matter 
is solely triggered by the underlying quark-gluon dynamics. 

By lowering the RG scale, 
starting from~$k=\Lambda$, the quark-gluon vertex generates four-quark self-interactions via two-gluon exchange. With respect to this type of interaction channels, 
we only take into account the diquark channel as discussed above. This channel is associated with the coupling~$\bar{\lambda}_\text{csc}$ in Eq.~\eqref{eq:gammak}. 
Once generated, this four-quark interaction channel can be removed by mapping it onto a Yukawa-type quark-diquark interaction channel associated 
with the coupling~$\bar{h}$ and a term bilinear in the diquark fields associated with the term $\sim\bar{m}^2$. Essentially, this corresponds to performing 
a Hubbard-Stratonovich transformation at a given RG scale. In the next RG step, however, 
the four-quark interaction~$\bar{\lambda}_\text{csc}$ is regenerated by the quark-gluon vertex and the quark-diquark vertex. The regenerated 
four-quark channel can then again be removed by mapping it onto the quark-diquark interaction channel and the term 
bilinear in the diquark fields. Moreover, the running of the quark-diquark coupling~$\bar{h}$ and the parameter~$\bar{m}^2$ receive additional contributions 
from, e.g., the running of the wavefunction renormalization~$Z_{\Delta}$ of the diquark fields. The latter is generated itself by the 
quark-diquark coupling~$\bar{h}$. Note that, once the diquark wavefunction renormalization is rendered finite, the diquarks become dynamical 
degrees of freedom in the RG flow. It is also important to add that higher-order diquark self-interaction terms are generated 
via the aforementioned quark-diquark interactions. In the following, we take into account diquark self-interactions up to the four-diquark channel which is 
associated with the coupling~$\bar{\lambda}_{\Delta}$ in Eq.~\eqref{eq:gammak}.\footnote{Such diquark self-interaction channels can be related to  
higher-order quark self-interaction channels with a nontrivial momentum structure. For example, the four-diquark channel can be related to an 
eight-quark interaction channel.} Taking another RG step, the four-quark interaction channel is then generated again and the aforementioned procedure 
of mapping it onto a Yukawa-type quark-diquark interaction channel and a term 
bilinear in the diquark fields can be repeated. The repeated application of this mapping can be recast into flow equations which 
eventually allow us to follow the RG flow from the perturbative high-momentum regime governed by quark-gluon dynamics 
down to  the low-energy regime governed by the formation of bound states of quarks. Within the functional 
RG framework, this procedure can be implemented with the aid of the so-called dynamical hadronization 
technique~\cite{Gies:2001nw,Gies:2002hq,Pawlowski:2005xe,Gies:2006wv,Floerchinger:2009uf,Braun:2014ata,Fu:2019hdw,Fukushima:2021ctq}, as already indicated  
in the previous subsection.

Employing this technique, see App.~\ref{app:dynhad} for details, we find 
the following coupled set of flow equations for 
the dimensionless renormalized
curvature of the effective potential,~$\epsilon_{\mu}=(\bar{m}^2 - 4 Z_\Delta \mu^2)Z_\Delta^{-1}k^{-2}$,
the renormalized four-diquark coupling~$\lambda_\Delta=\bar{\lambda}_\Delta Z_\Delta^{-2}$, and
the renormalized quark-diquark coupling~$h= \bar{h} Z_\Delta^{-\frac{1}{2}}$:
{\allowdisplaybreaks
\be
\partial_t\epsilon_{\mu} &=& (\eta_{\Delta}-2)\epsilon_{\mu} -8  h^2 b_{(2,0)}(\tilde{\mu},0) \nn\\ 
&& \qquad +\frac{2}{h^2}\epsilon_{\mu}\left(1+\epsilon_{\mu}\right) g^4  b_{(0,4)}^{(A)}(\tilde{\mu},0,\eta_A)\,,
\label{eq:eflow}
 \\
\partial_t\lambda_\Delta&= & 2\eta_\Delta\lambda_\Delta + 
4 h^4 b_{(4,0)}(\tilde{\mu},0) \nn\\
&& \qquad +4\frac{\lambda_\Delta}{h^2} \left( 1 + \epsilon_{\mu}\right) g^4b_{(0,4)}^{(A)}(\tilde{\mu},0,\eta_A)\,,
\label{eq:lflow}
\\
\partial_t h^2&=&\eta_\Delta h^2 + 
 \frac{16}{3}g^2h^2 b_{(1,2)}^{(A)}(\tilde{\mu},0,\eta_A) \nn\\ 
 &&  \qquad + 2 \left( 1+2\epsilon_{\mu}\right) g^4  b_{(0,4)}^{(A)}(\tilde{\mu},0,\eta_A)\,,
 \label{eq:hflow}
\ee
where~$\tilde{\mu}=\mu/k$ is the dimensionless chemical potential 
and~$g^2=\bar{g}^2 Z_A^{-1}$ is the renormalized strong coupling. 
Finally, the scale-dependence of the anomalous dimension of the diquark fields is governed by
\be
\eta_\Delta=-\partial_t \ln Z_\Delta =8 h^2 d_{(2,0)}(\tilde{\mu},0)\,.
\label{eq:etaDelta}
\ee
Recall that the set of couplings associated with these equations span our ansatz~\eqref{eq:gammak} for the scale-dependent effective action~$\Gamma_k$.
The functions~$b_{(i,j)}$ and~$d_{(i,j)}$ are so-called threshold functions which correspond to one-particle irreducible (1PI) Feynman diagrams with~$i$ 
external bosonic and~$j$ external fermionic lines, respectively. In some cases, an additional superscript~$(A)$ appears which indicates that 
the associated diagram contains at least one gluon line. Since we restrict ourselves 
to the zero-temperature limit in this work, these functions only depend on the dimensionless chemical potential $\tilde{\mu}$ in the absence of a diquark gap. 
In any case, the regularization scheme dependence is also encoded in these functions. In this respect, we note that we employ a scheme which allows us 
to integrate out fermionic fluctuations around the Fermi surface~\cite{Braun:2020bhy}, see  App.~\ref{app:tfcts} for 
its definition and brief discussion of all threshold functions entering our present study.
It should be emphasized that also the anomalous dimensions~$\eta_{\Delta}$ and~$\eta_A=-\partial_t \ln Z_A$ depend on the dimensionless 
chemical potential~$\tilde{\mu}$. The running of the strong coupling~$g$ and its relation to the wavefunction renormalization~$Z_A$ of the gauge fields 
is discussed below.

By comparing the set of flow equations~\eqref{eq:eflow}-\eqref{eq:etaDelta} with our ansatz~\eqref{eq:gammak} for the effective action, it becomes apparent that 
there is no flow equation for the four-quark coupling~$\bar{\lambda}_\text{csc}$. With the aid of the aforementioned dynamical hadronization technique~\cite{Gies:2001nw,Gies:2002hq,Pawlowski:2005xe,Gies:2006wv,Floerchinger:2009uf,Braun:2014ata,Fu:2019hdw},
the contributions to this coupling are continuously transformed into contributions to the flow of the quark-diquark coupling~$h$ and the 
curvature~$\epsilon_{\mu}$, such that~$\partial_t \bar{\lambda}_\text{csc} =0$ for any value of~$k$. The contributions to the flow 
of the four-quark coupling~$\bar{\lambda}_\text{csc}$ therefore appear in the flow equations for the quark-diquark coupling~$h$ and the 
curvature~$\epsilon_{\mu}$. In particular, these contributions are associated with the terms~$\sim g^4$ in the flow equations~\eqref{eq:eflow} 
and~\eqref{eq:hflow} which originally stem from two-gluon exchange box diagrams 
appearing in the RG flow of four-quark couplings.\footnote{It is indeed possible to recover the flow equation for the four-quark coupling~$\bar{\lambda}_\text{csc}$  
at scales above the symmetry breaking scale~$k_{\text{SB}}$. 
To be more specific, we have~${\lambda}_\text{csc} = h^2/(2\epsilon_{\mu})$ for the dimensionless four-quark coupling which relates  
the dimensionless four-quark coupling to the quark-diquark coupling~$h$ and the curvature~$\epsilon_{\mu}$.
From this flow equation, we deduce that the RG flow at sufficiently large 
scales~$k$ is governed by the two fixed points of~$\lambda_\text{csc}$, provided the strong coupling~$g^2$ is sufficiently small, 
see Subsec.~\ref{subsec:dg}. These fixed points 
can be translated into fixed points for the quark-diquark coupling~$h$ and the curvature~$\epsilon_{\mu}$, see Ref.~\cite{Braun:2011pp} for a general discussion 
of this aspect.}

The set of flow equations~\eqref{eq:eflow}-\eqref{eq:etaDelta} describes the dynamics at high-momentum scales where the curvature~$\epsilon_{\mu}$ of the effective potential 
is positive. In fact, as discussed above, we shall choose initial conditions such that~$\epsilon_{\mu}\gg 1$,~$\lambda_\Delta\to 0$, and~$h^2\to 0$
for~$k\to \Lambda$. Quark self-interactions, which are mapped onto diquark self-interactions and quark-diquark interactions 
in our present setting, are initially only generated by 
two-gluon exchange~$\sim g^4$. Following the RG flow to smaller scales~$k$, we find that 
the curvature~$\epsilon_{\mu}$ decreases and eventually becomes zero at a finite scale~$k_{\text{SB}}$, 
see also our discussion in Sec.~\ref{sec:RGhighdens} below. At the scale~$k_{\text{SB}}$, spontaneous~${\rm U}(1)_{\text{V}}$ symmetry
breaking sets in. 

Below the symmetry breaking scale~$k_{\text{SB}}$, the curvature~$\epsilon_{\mu}$ of the effective potential becomes negative and 
a color-superconducting ground state is formed associated with the formation of a gap in the fermionic excitation spectrum. Note that 
the antisymmetric flavor structure of this color-superconducting ground state corresponds to a singlet representation of the global chiral group. 
This implies that the formation of such a ground state does not violate the chiral symmetry. In any case, 
for~$k\leq k_{\text{SB}}$, it is convenient to switch from the set of flow equations~\eqref{eq:eflow}-\eqref{eq:hflow} 
to a set in which the flow equation for the curvature~$\epsilon_{\mu}$ is replaced with a flow equation for the 
minimum~$|\Delta_0|^2= \sum_a |\Delta_{0,a}|^2$. 
Recall that we expand the effective action in the quantity~$\Delta_a^{\ast}\Delta_a$ (summation over~$a$ is assumed).
For convenience, we shall choose~$\Delta_{0,a} = \Delta_0\delta_{a,3}$ ($\Delta_0 \in {\mathbb R})$
and use~$\kappa = Z_{\Delta}|\Delta_0|^2 k^{-2}$ to parametrize the flow of the position of the minimum of the 
effective action. The resulting set of flow equations for scales~$k < k_{\text{SB}}$ then reads 
{\allowdisplaybreaks
\be
\partial_t\kappa &=& -(\eta_{\Delta} + 2)\kappa +  \frac{4 h^2   }{ \lambda_\Delta  } b_{(2,0)}(\tilde{\mu},h^2\kappa)\,,\\
\partial_t\lambda_\Delta&=&  2\eta_\Delta\lambda_\Delta + 
4 h^4    b_{(4,0)}\big(\tilde{\mu},h^2\kappa\big) \nn \\ 
&& \qquad\qquad\qquad  +4\frac{\lambda_\Delta}{h^2} g^4  b_{(0,4)}^{(A)}\big(\tilde{\mu},h^2\kappa,\eta_A\big)\,,\\
\partial_t h^2&=& \eta_\Delta h^2 +  \frac{16}{3}g^2 h^2   b_{(1,2)}^{(A)}\big(\tilde{\mu},h^2\kappa,\eta_A\big) \nn\\ 
&& \qquad\qquad\qquad +2 g^4  b_{(0,4)}^{(A)}\big(\tilde{\mu},h^2\kappa,\eta_A\big)\,,
\label{eq:hbroken}
\ee
and}
\be
\eta_\Delta= 8 h^2 d_{(2,0)}(\tilde{\mu},h^2\kappa)\,.
\label{eq:etabroken}
\ee
In addition to the dimensionless chemical potential~$\tilde{\mu}$, the anomalous dimensions~$\eta_{\Delta}$ and~$\eta_A$ now 
also depend on the so-called (diquark) gap~$|\Delta_{\text{gap}}|$ which appears in the propagator of the quarks. 
In our conventions, we have~$|\Delta_{\text{gap}}| = h\sqrt{\kappa}k$. Thus, the gap~$|\Delta_{\text{gap}}|$ 
in the quark propagator is directly related to the minimum~$|\Delta_0|^2$.

 The gauge sector enters our flow equations~\eqref{eq:eflow}-\eqref{eq:etabroken} only via the running of the strong coupling~$g$ 
 which is governed by the following equation:
 \be
 \partial_t g^2 = \eta_A g^2\,.
 \label{eq:g2flow}
 \ee
 Here,~$\eta_A$ can be decomposed into a pure gluonic contribution~$\eta_{\text{glue}}$ and a term~$\eta_q$ which contains  
 the quark contributions~\cite{Gies:2005as, Braun:2005uj, Braun:2006jd}: 
 \be
 \eta_A = -\partial_t \ln Z_A = \eta_{\text{glue}} + \eta_{q}\,.
 \label{eq:ZAflowmt}
 \ee
 For the purely gluonic contribution~$\eta_{\text{glue}}$, we employ the results from previous functional RG studies~\cite{Gies:2002af, Braun:2005uj, Braun:2006jd}. 
 There,~$\eta_{\text{glue}}$ has been computed non-perturbatively within the background field formalism which also underlies our present work. 
 The quark contribution~$\eta_q$ depends on the dimensionless chemical potential~$\tilde{\mu}$ and the diquark gap:
 \be
 \eta_q = \frac{1}{4}g^2 d^{(A)}_{(2,0)}(\tilde{\mu}, h^2\kappa)\,. 
 \label{eq:etaq_gauge}
 \ee
In the limit~$\tilde{\mu}\to 0$ and~$h^2\kappa\to 0$, we have $d^{(A)}_{(2,0)}(0,0) = 2/(3\pi^2)$ and therefore~$\eta_q = g^2/(6\pi^2)$. This is nothing 
but the standard one-loop contribution of the quark fields to the running of the strong coupling~$g$. 
We add that, in general, the running of the gauge coupling also receives corrections from quark self-interactions, such as four-quark interactions, 
see, e.g., Refs.~\cite{Gies:2003dp,Kusafuka:2011fd}. However, within 
the fRG framework, it follows from an analysis of (modified) Ward-Takahashi identities  that 
such back-reactions of the matter sector on the gauge sector are negligible, provided that the flow of the four-quark couplings is governed by 
the presence of fixed points~\cite{Gies:2005as, Braun:2005uj, Braun:2006jd}. At least above the symmetry breaking scale~$k_{\text{SB}}$, this is indeed the 
case in our present study (see also Subsec.~\ref{subsec:sf}) which justifies that we do not take such contributions to the running of the gauge 
sector into account, see also Ref.~\cite{Braun:2019aow}. For~$k<k_{\text{SB}}$, we shall neglect such contributions. Note that, in this regime, the situation 
is particularly involved anyhow because of the presence of a finite quark gap, as we shall discuss next and also in Sec.~\ref{sec:RGhighdens} below. 

In our flow equations~\eqref{eq:eflow}-\eqref{eq:etaq_gauge} we drop fluctuations of the diquark fields. Such fluctuation effects are associated 
with 1PI diagrams coming with at least one internal diquark line. Compared to the contributions that 
we take into account in our analysis, such contributions are subleading in an $N_{\text c}$-counting. Moreover, in the symmetric high-energy regime 
(i.e., for~$k>k_{\text{SB}}$), the fluctuation effects of the diquark fields are parametrically suppressed because of the large diquark mass 
parameter. In Sec.~\ref{sec:RGhighdens}, we shall see that this parameter is indeed large and only becomes small close to the symmetry breaking scale~$k_{\text{SB}}$. 
Such a suppression of fluctuation effects has already been observed and discussed  in 
early fRG studies of chiral models in the zero-density limit~\cite{Jungnickel:1995fp,Berges:1997eu,Berges:2000ew}. 

In the regime~$k<k_{\text{SB}}$, which is governed by spontaneous symmetry breaking, it can no longer be argued that fluctuation effects 
are subleading. Whereas fluctuation effects are associated with, e.g., pion dynamics
at low densities, a rigorous inclusion of fluctuations of the diquark fields at high densities requires to deal with 
an Anderson-Higgs-type mechanism~\cite{PhysRev.130.439,Englert:1964et,Higgs:1964ia,Higgs:1964pj,Guralnik:1964eu} 
associated with the symmetry-breaking pattern~$\text{SU}(3)\to \text{SU}(2)$ 
in color space (as the diquark fields carry a net color charge). 
As a consequence, only three of the eight gluons are massless. The remaining five gluons are effectively 
rendered massive by, loosely speaking, 
``eating up" Goldstone modes which appear in the diquark spectrum in the symmetry-broken regime, see, e.g., Ref.~\cite{Alford:2007xm} for a review.
In our present study, which mainly aims at setting the methodological stage for future more quantitative studies of dense QCD matter, 
we do not include this Anderson-Higgs-type mechanism but rather drop diquark fluctuations as mentioned above. 
A more quantitative study taking this Anderson-Higgs-type mechanism into account is deferred to future work. The general methodological 
groundwork for studies of this type of mechanism within the fRG framework has already been laid in studies of Abelian Higgs models~\cite{Reuter:1992uk,Litim:1995kn} 
and (non-Abelian) gauged chiral Higgs-Yukawa models~\cite{Gies:2013pma}.
In any case, we shall at least 
estimate the effect of the appearance of the associated gap for the gluons on our present results in Sec.~\ref{sec:RGhighdens} below. 
\begin{figure*}[t]
  \includegraphics[width=\columnwidth]{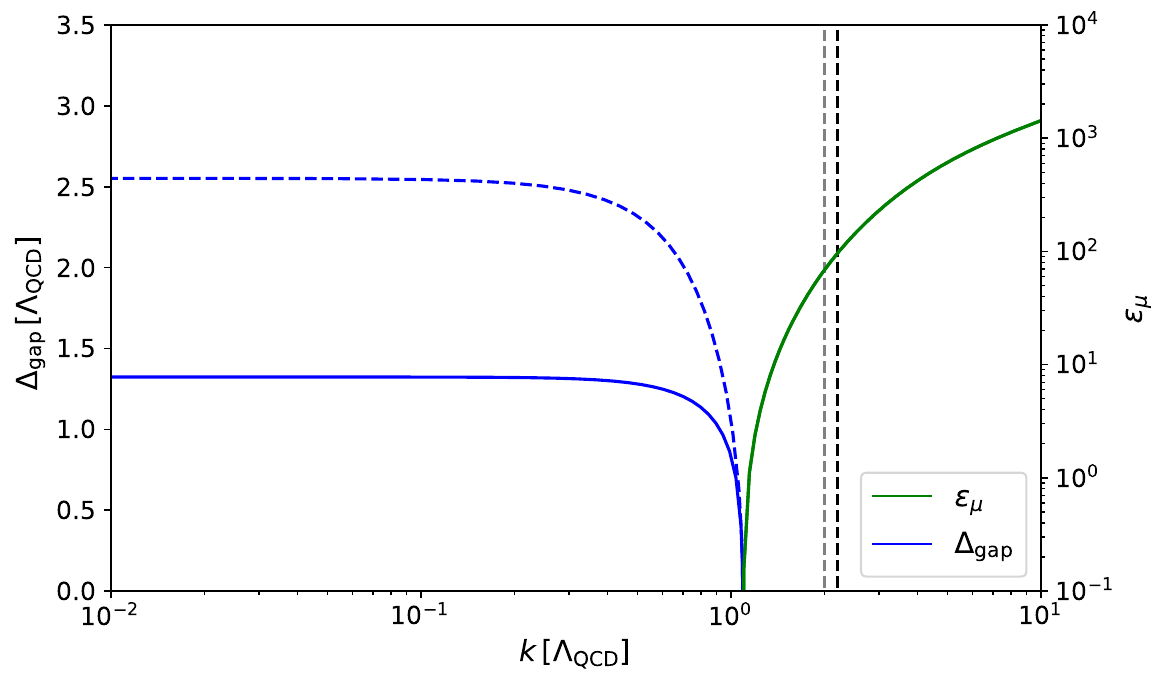}\hfill
  \includegraphics[width=0.92\columnwidth]{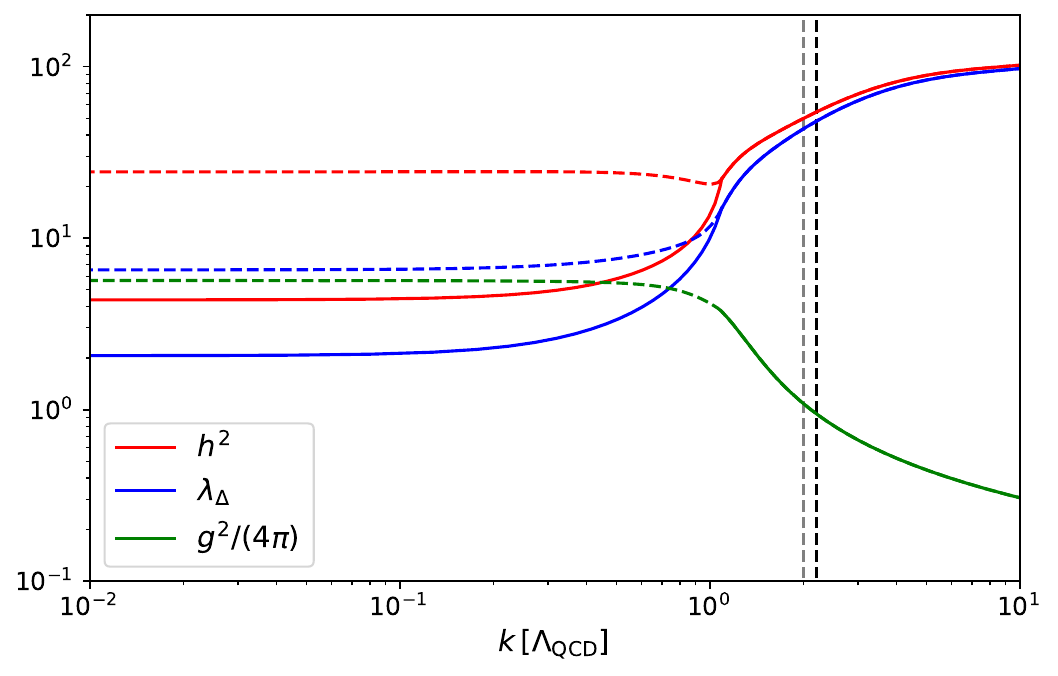}
\caption{RG flow of the renormalized dimensionless curvature~$\epsilon_{\mu}$ for~$k\geq k_{\text{SB}}$ (left panel), 
the diquark gap~$\Delta_{\text{gap}}$ for~$k\leq k_{\text{SB}}$ (left panel), the (squared) renormalized quark-diquark coupling~$h^2$ (right panel), the 
renormalized four-diquark coupling~$\lambda_{\Delta}$ (right panel), and the renormalized strong coupling~$\alpha = g^2/(4\pi)$ (right panel)
for~$\mu/\LQCD = 2$, where~$\ksb/\LQCD \approx 1.09$. In both panels, the gray (vertical) dashed line is associated with the scale~$k=\mu$. The black (vertical) dashed 
line in these panels is associated with the scale~$k=k_{\text{m}}$. Here, $k_{\text{m}}$ is an estimate for the scale at which the gluon 
screening masses exceed the scale~$k$. 
For~$k<\ksb$, the results for the diquark gap~$\Delta_{\text{gap}}$ (left panel) and the couplings in the right panel are given as solid and 
dashed lines. The dashed lines represent the running of these quantities 
for the case in which the gluons remain ungapped and do not acquire a mass according to the Anderson-Higgs mechanism below the symmetry breaking scale~$k_{\text{SB}}$. 
The solid lines show the results for the case in which the gluons have been fully decoupled from the matter sector for~$k\leq k_{\text{SB}}$.
}
\label{fig:flow}
\end{figure*}
\section{RG flow of dense QCD matter}
\label{sec:RGhighdens}
\subsection{Scale fixing}\label{subsec:sf}
Let us now discuss our results for the RG flow of dense QCD matter, in particular those for the chirally symmetric (scalar) diquark condensate. 
To this end, we first need to specify the initial conditions of our RG flow equations at the UV scale~$k=\Lambda=10\,\text{GeV}$. 
This value of the initial scale ensures that we have~$\Lambda\gg \mu$ for all values of the quark chemical potential considered in the present work. 
For the dimensionless renormalized curvature~$\epsilon_{\mu}$ of the effective potential, 
we choose~$\epsilon_{\mu} = 10^6$. Thus, the diquark fields do not represent dynamical degrees of freedom at the UV scale~$\Lambda$. 
We add that the limit~$\epsilon_{\mu}\to\infty$ corresponds to the limit 
of a vanishing diquark wavefunction renormalization,~$Z_{\Delta}\to 0$. 

For the quark-diquark coupling~$h$, we choose~$h=0.1$ at~$k=\Lambda$. The initial value of the four-diquark coupling~$\lambda_{\Delta}$ is set 
to zero. This choice for the couplings at the scale~$\Lambda$ ensures that we indeed initialize the flow ``in the vicinity of"
the QCD action in the UV limit.\footnote{We add that a finite value of the quark-diquark coupling~$h$ explicitly 
breaks the~${\rm U}(1)_{\rm A}$ symmetry. As discussed in, e.g., Ref.~\cite{Braun:2019aow}, 
this is required to render 
the four-quark coupling~$\lambda_{\text{csc}}$ associated with the diquark 
channel~$\sim (\psib_b\tau_2\epsilon_{abc}\gamma_5\CC \psib^T_c)(\psi^T_d\CC\gamma_5\tau_2\epsilon_{ade}\psi_e)$ 
to be most dominant at high densities~\cite{Braun:2019aow}.}
Note that we have checked that our results in the IR limit (in particular those for the diquark gap) 
depend only very weakly on the specific choice for the initial values of the couplings, provided that we ensure~$\Lambda \gg k_{\text{SB}}$. 
This independence can be traced back to the appearance of pseudo fixed points in the RG flow of the gluon-induced interaction channels,\footnote{In the present study, we only encounter pseudo fixed points 
since a dimensionful scale enters the RG flow via the quark chemical potential.} see Refs.~\cite{Gies:2002hq,Gies:2006wv,Floerchinger:2009uf,Braun:2014ata,Fu:2019hdw} for a 
detailed discussion 
of this aspect in the vacuum limit of QCD. The appearance of a pseudo fixed-point behavior at finite chemical potential 
together with a loss of memory of the details of the initial conditions may 
already be anticipated from an analysis of the fixed-point structure of gluon-induced four-quark interaction channels, see Ref.~\cite{Braun:2019aow} for details. 
Indeed, the quark-diquark coupling and the curvature of the effective potential can be directly related to the four-quark 
coupling~${\lambda}_\text{csc} = h^2/(2\epsilon_{\mu})$. Therefore, fixed points of the four-quark coupling~${\lambda}_\text{csc}$ leave their imprint 
in the RG flows of the quark-diquark coupling~$h$ and the curvature~$\epsilon_{\mu}$. For example, 
our choice~$\epsilon_{\mu}\gg h^2$ at the UV scale implies that we initialize the RG flow (very) close to the Gau\ss ian 
fixed point of the four-quark coupling~$\lambda_{\text{csc}}$. 

From this discussion it follows that the initial value of the strong coupling is the only input parameter in our calculations. 
It sets the scale for all dimensionful 
quantities. In our present study with two massless quark flavors, 
we choose~$\alpha=g^2/(4\pi) = 0.179\pm 0.004$ at the UV scale~$\Lambda=10\,\text{GeV}$ which corresponds 
to the experimental value~$\alpha = 0.330\pm 0.014$ at the $\tau$-mass scale~\cite{Bethke:2009jm}.\footnote{Note that the running of the strong coupling entering our 
calculation is compatible with the standard $\overline{\text{MS}}$ running over a wide range of scales~\cite{Gies:2002af,Braun:2005uj,Braun:2006jd}.} 
For~$\Lambda_{\text{QCD}}$ -- defined as the inflection point of the strong coupling -- we then obtain~$\Lambda_{\text{QCD}} \approx 209\,\text{MeV}$ in the vacuum limit. From here on, we shall 
measure all dimensionful quantities in units of~$\Lambda_{\text{QCD}}$. For example, we have $\Lambda/\Lambda_{\text{QCD}}\approx 47.8$.

\subsection{From quark-gluon dynamics to color superconductivity}
In Fig.~\ref{fig:flow}, we show the RG flow of the renormalized dimensionless curvature~$\epsilon_{\mu}$ for~$k\geq k_{\text{SB}}$ (left panel), 
the diquark gap~$\Delta_{\text{gap}}$ for~$k\leq k_{\text{SB}}$ (left panel), the (squared) quark-diquark coupling~$h^2$ (right panel), 
the four-diquark coupling~$\lambda_{\Delta}$ (right panel), and the strong coupling~$\alpha=g^2/(4\pi)$ (right panel) 
over a wide range of 
scales for~$\mu/\LQCD = 2$. In this case, we have~$\ksb/\LQCD \approx 1.09$. The gray (vertical) dashed lines in 
the two panels represent the point in the RG flow where~$k=\mu$. The black (vertical) dashed 
lines are associated with the scale~$k=k_{\text{m}}$. The latter is an estimate for the scale at which the screening 
masses of the gluons exceed the scale~$k$.\footnote{For simplicity, we do not distinguish between the electric and magnetic masses.} 
Here, we estimate this scale from the relation~$m_{\text{g}}= g(k_{\text{m}})\mu/\pi = k_{\text{m}}$, 
where~$m_{\text{g}}$ represents an estimate for the gluon screening masses  
in the symmetric high-energy regime ($k>\ksb$), see, e.g., Refs.~\cite{Kapusta:1979ux,Toimela:1984xy}. Note that these masses are scheme-dependent 
quantities. A detailed analysis of this aspect will be given elsewhere~\cite{BGS}. In any case, for small chemical potentials (e.g., $\mu/\LQCD=2$ 
as shown in Fig.~\ref{fig:flow}), we observe a hierarchy of scales: $\ksb < \mu < k_{\text{m}}$. 

For~$k> k_{\text{m}}$, gluon screening effects are parametrically suppressed since~$m_{\text{g}}/k < 1$. Note that effects associated with the 
quark chemical potential appearing in the quark propagator are even more suppressed,  $\mu/k < m_{\text{g}}/k < 1$. 
In this high-energy regime, we therefore do not expect that our results 
suffer significantly from the fact that we have neglected the gluon screening masses in our calculations. 
For these scales, the RG flow of the couplings is mainly driven by gluon exchange diagrams. 
Following the RG flow towards smaller scales, the strong coupling increases (see right panel of Fig.~\ref{fig:flow}) and gauge fluctuations tend to drive the system towards 
a ground state associated with a (spontaneously) broken ${\rm U}(1)_{\rm V}$ symmetry. However, it should be noted that {\it strong} gauge fluctuations are in principle not required to 
trigger the formation of a (color-)superconducting ground state because of the presence of a Cooper instability in the system,\footnote{This is different for chiral symmetry breaking 
which requires the gauge coupling to become sufficiently large, see, e.g., Refs.~\cite{Gies:2005as,Braun:2005uj,Braun:2006jd,Braun:2011pp,Braun:2019aow} for a detailed discussion.} 
see Refs.~\cite{Braun:2017srn,Braun:2018bik} for a general fixed-point analysis of this aspect and Ref.~\cite{Alford:1997zt} for an early mean-field study in QCD.
The gauge fluctuations rather act as a ``catalyzer" for the formation of a (color-)superconducting ground state. Loosely speaking, 
strong gauge fluctuations tend to increase the symmetry breaking scale~$\ksb$ and therefore also the diquark gap~$\Delta_{\text{gap}}\sim \ksb$. In other words, without 
strong gauge fluctuations, the diquark gap would be (significantly) smaller.

From this line of arguments it is already clear that gluon screening effects become relevant at some point in the RG flow towards the infrared regime. To be more specific, 
we expect that the presence of gluon screening masses affects the dynamics for~$\ksb < k < k_{\text{m}}$. In this regime, contributions to the RG flow with at least one internal gluon 
line start to become parametrically suppressed since we have $m_{\text{g}}/k >1$. This reduces the aforementioned ``catalyzing effect" of the gluons 
and presumably leads to a shift of the symmetry breaking scale~$\ksb$ 
and the gap~$\Delta_{\text{gap}}$ 
to smaller values compared to the ones obtained in our present study. A detailed analysis of this aspect will be given elsewhere.
In any case, we expect that 
their inclusion will not significantly alter the value of the symmetry breaking scale~$\ksb$ or 
the dynamics for~$\ksb < k < k_{\text{m}}$, at least for sufficiently small values of the chemical potential. 
For example, for~$\mu/\LQCD =2$, we have~$\ksb/\LQCD \approx 1.1$ 
and~$k_{\text{m}}/\LQCD \approx 2.2$. Consequently, 
gluon screening effects are expected to be relevant only in a comparatively small regime above the symmetry breaking scale. 

For increasing chemical potential, we find that the symmetry breaking scale~$\ksb$ increases but only mildly, see also Fig.~\ref{fig:gap} and 
our discussion in Subsec.~\ref{subsec:dg} below. 
In any case, a change in the hierarchy of scales sets in for increasing~$\mu$, where we eventually have~$\ksb < k_{\text{m}} < \mu$. 
The quark dynamics is now strongly affected by the presence of the 
chemical potential over a wide range of scales. 
For~$\mu/\LQCD = 10$, for example, we have~$\ksb/\LQCD\approx 1.6$. The gluon screening masses are smaller than the quark chemical potential 
over a wide range of scales within the regime~$\ksb < k_{\text{m}} < \mu$.
Nevertheless, these screening masses increase roughly linearly when~$\mu$ is increased. 
For a given scale~$k$, this suggests a stronger (parametric) suppression of gluonic contributions to the RG flow 
at large chemical potential than at small chemical potential. In other words, 
gluon screening effects may have a stronger impact on the RG flow over a wider range of scales when the chemical potential is increased. 
Correspondingly, the aforementioned ``catalyzing effect" of the 
gauge degrees of freedom is expected to be reduced. 
Therefore, it is reasonable to expect that our estimates for the symmetry breaking scale~$\ksb$ and the diquark gap 
become less reliable for large chemical potentials. 
In fact, after the conventional BCS-type increase of~$\ksb$ for small chemical potentials, 
this suggests that gluon screening effects may potentially even lead to a decrease of~$\ksb$ 
over some range of quark chemical potentials. For chemical potentials beyond those considered in this work, however, 
it is known that the diquark gap increases again as a function of the chemical potential~\cite{Son:1998uk}, 
see also Refs.~\cite{Hong:1999fh,Schafer:1999jg,Pisarski:1999bf,Hsu:1999mp,Brown:1999aq,Schafer:1999fe} for a discussion 
of the relevance of gluon screening effects.

Let us now turn to the regime~$k<\ksb$ associated with spontaneous ${\rm U}(1)_{\rm V}$ breaking. 
In this regime, the situation is even more involved as it requires to deal with 
an Anderson-Higgs-type mechanism~\cite{PhysRev.130.439,Englert:1964et,Higgs:1964ia,Higgs:1964pj,Guralnik:1964eu} associated with the 
breaking of the~$\text{SU}(3)$ symmetry in color space down to a $\text{SU}(2)$ symmetry. 
This eventually leads to the generation of ``gaps" (screening masses) for five of the eight gluons. A rigorous treatment of this mechanism 
is beyond the scope of the present work. We only consider two approximations in the low-energy regime~$k<\ksb$ to already gain some understanding of the effect of 
gluon screening in the long-range limit. 
\begin{figure*}[t]
  \includegraphics[width=1\columnwidth]{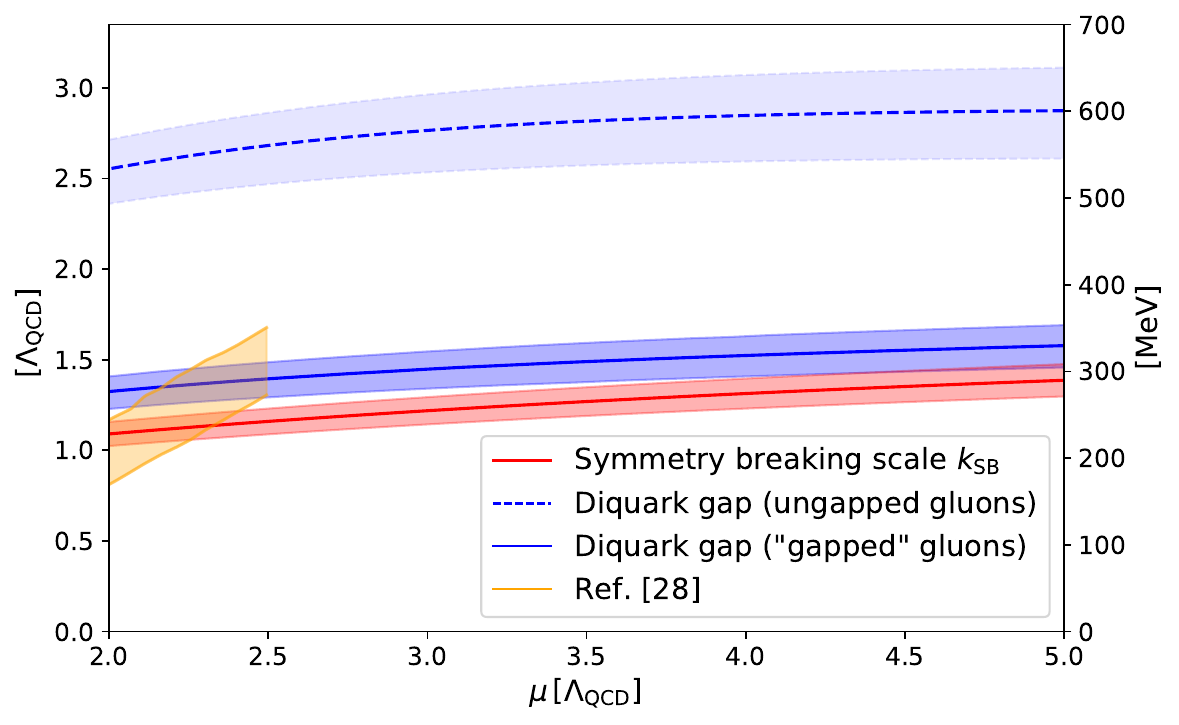}\hfill
    \includegraphics[width=0.925\columnwidth]{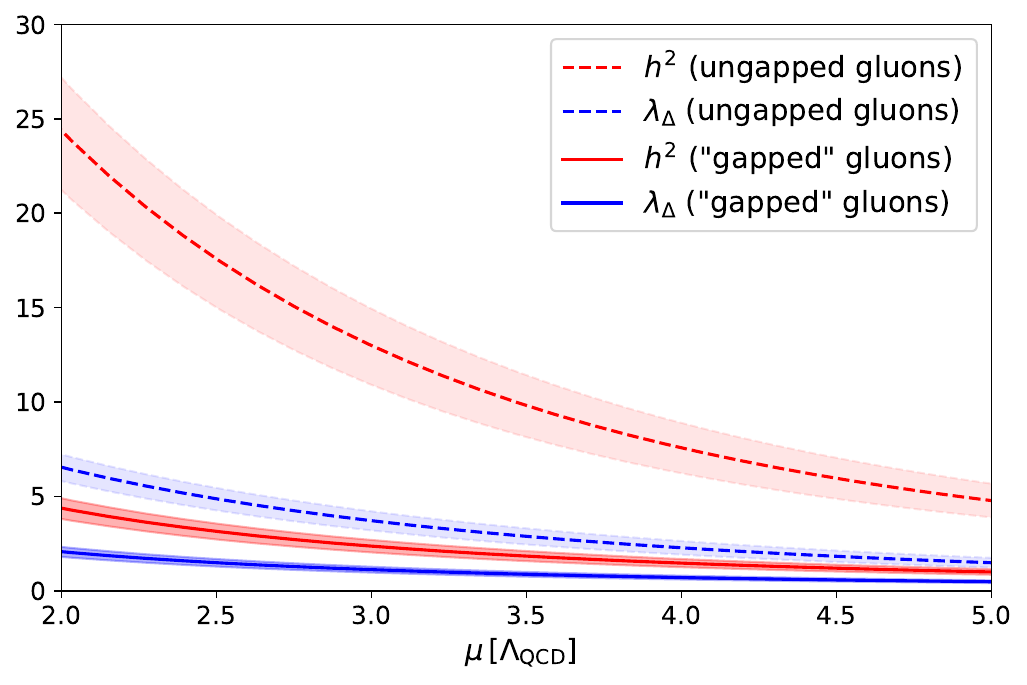}
\caption{Left panel: Diquark gap~$\Delta_{\text{gap}}$ and the symmetry-breaking scale~$\ksb$ 
as a function of the quark chemical potential~$\mu$. 
The shaded bands (apart from the orange band) represent the uncertainty resulting from a variation of the strong coupling at the initial RG scale. 
The blue dashed line together with the light blue band (ungapped gluons) represent the gap 
for the case where the gluons remain ungapped below the symmetry breaking scale~$k_{\text{SB}}$. The solid blue line together with the 
dark blue band (``gapped" gluons) show the gap for the case where the gluons have been decoupled from the matter sector for~$k\leq k_{\text{SB}}$, 
see main text for details. The orange band depicts results for the diquark gap 
from a previous fRG study~\cite{Leonhardt:2019fua}. Note that, in Ref.~\cite{Leonhardt:2019fua}, the shown 
range of chemical potentials is associated with densities~$n/n_0 \approx 6\dots 12$ (where~$n_0$ is the nuclear saturation density). 
The results from Ref.~\cite{Leonhardt:2019fua} are in remarkable agreement with those from early studies of the diquark gap for~$n/n_0\lesssim 5$~\cite{Alford:1997zt} . 
For $n/n_0\approx 5$, for example,~$\Delta_{\text{gap}}\approx 70\dots 160\,\text{MeV}$ was reported in Ref.~\cite{Alford:1997zt} 
and~$\Delta_{\text{gap}}\approx 140\dots 230\,\text{MeV}$ was found in Ref.~\cite{Leonhardt:2019fua}.
Right panel: The (squared) renormalized quark-diquark coupling~$h^2$ and the 
renormalized four-diquark coupling~$\lambda_{\Delta}$ as a function of the quark chemical potential~$\mu$. 
Dashed and solid lines are again associated with ungapped and ``gapped" gluons in the low-energy regime, respectively. 
The shaded bands represent the uncertainty resulting from a variation of the strong coupling at the initial RG scale. 
}
\label{fig:gap}
\end{figure*}

In the first approximation, we simply leave the gluons ungapped for~$k<\ksb$. The corresponding results for the
RG flow of the diquark gap~$\Delta_{\text{gap}}$ and the various couplings are depicted by the dashed lines in Fig.~\ref{fig:flow} (and also in Fig.~\ref{fig:gap}). 
In the second approximation 
associated with the solid lines for~$k<\ksb$ in Fig.~\ref{fig:flow} (and also in Fig.~\ref{fig:gap}), we decouple the gluon contributions from the RG flow of the matter sector, which may be viewed as  
adding an ``infinite gap" to all gluons. In practice, we have implemented this decoupling by setting the 
gauge coupling to zero for~$k<\ksb$. Comparing the corresponding results with the ones for the ungapped gluons, we observe that the diquark gap 
is reduced by roughly a factor of two. As already discussed above, ungapped/unscreened gluons indeed act as a ``catalyzer" for the formation of a \mbox{(color-)} superconducting ground state. 
This remains also true when the chemical potential is increased, see Fig.~\ref{fig:gap}. For the renormalized quark-diquark coupling~$h$ and the 
renormalized four-diquark coupling~$\lambda_{\Delta}$, we observe a similar behavior. The couplings receive a significant boost in the approximation with ungapped gluons, 
see the right panels of Figs.~\ref{fig:flow} and~\ref{fig:gap}.

\subsection{Diquark gap}\label{subsec:dg}
 Our results for the diquark gap~$\Delta_{\text{gap}}$, the symmetry breaking scale~$\ksb$, the quark-diquark coupling~$h$, and the four-diquark coupling~$\lambda_{\Delta}$ in 
the limit~$k\to 0$ 
as a function of the quark chemical potential are summarized in Fig.~\ref{fig:gap}. The (shaded) bands in Fig.~\ref{fig:gap} 
result from a variation of the strong coupling at the initial RG scale, see Subsec.~\ref{subsec:sf}.  Note that the variation of our results 
arising from a variation of the regularization scheme (associated with regulator functions) is negligible compared to the one obtained from the aforementioned 
variation of the initial value of the strong coupling. We refer the reader to App.~\ref{app:tfcts} for the definition of the regulator functions  
employed in the present work and a corresponding discussion of the scheme dependence.

The observed dependence of the symmetry breaking scale~$\ksb$ on the chemical potential appears consistent with the standard 
BCS-type scaling behavior. This is true for the case with ungapped gluons in the low-energy regime and for the case with decoupled gluon contributions as
associated with (infinitely) ``gapped" gluons. However, the diquark gap in the case with ungapped gluons is found to be significantly greater than 
the one obtained in our calculations with ``gapped" gluons. 

Let us now analyze the scaling behavior of the symmetry breaking scale and the diquark gap in more detail. To this end, it is convenient to reconstruct the RG flow 
of the four-quark coupling~$\lambda_{\text{csc}}=h^2/(2\epsilon_{\mu})$ from the RG flows of the quark-diquark coupling~$h$ and the curvature~$\epsilon_{\mu}$ 
of the effective potential. Employing Eqs.~\eqref{eq:eflow} and~\eqref{eq:hflow}, we then find 
\be
\partial_t \lambda_{\text{csc}} &=& 2\lambda_{\text{csc}} + 16\lambda_{\text{csc}}^2 b_{(2,0)}\nn\\
&& \qquad\quad +\frac{16}{3} \lambda_{\text{csc}} g^2 b_{(1,2)}^{(A)}  + g^4 b_{(0,4)}^{(A)}
\,.
\label{eq:dtl0}
\ee
Setting~$\lambda_{\text{csc}}=0$ (corresponding to $\epsilon_{\mu}\gg h^2$) at the initial RG scale~$k=\Lambda\gg \mu$, the RG flow of~$\lambda_{\text{csc}}$ is then 
dominated by the contributions~$\sim g^4$ associated with two-gluon exchange diagrams. All the other contributions to the flow of this coupling are initially 
subleading. Thus, we are left with
\be
\partial_t \lambda_{\text{csc}} = 2\lambda_{\text{csc}} + g^4 b_{(0,4)}^{(A)}\,,
\label{eq:dtl1}
\ee
where~$b_{(0,4)}^{(A)}<0$ for~$\mu/k\to 0$, see App.~\ref{app:tfcts}. 

Let us now define a scale~$\bar{k}$ such that the dependence 
of the two-gluon exchange diagrams on the chemical potential is negligible for~$k>\bar{k}$. 
In this regime, the flow equation~\eqref{eq:dtl1} can be solved analytically. Integrating Eq.~\eqref{eq:dtl1} from~$k=\Lambda$ 
down to~$k=\bar{k}$, we find
\be
\lambda_{\text{csc}} (\bar{k}) = -\frac{1}{2}b_{(0,4)}^{(A)} g^{4}(\bar{k}) + {\mathcal O}(g^6)\,.
\label{eq:dtl2}
\ee
Here, we dropped terms which are subleading for~$\Lambda\gg \bar{k}$. 

We shall now also assume that~$\bar{k}$ can be chosen such that, at this scale, 
the gluon-induced four-quark self-interactions~$\sim\lambda_{\text{csc}}$ have become strong enough to ``dominate" their own 
RG flow. For sufficiently small values of the chemical potential~$\mu$, it may indeed be possible to choose~$\bar{k}$ such that 
the approximations underlying the derivations of Eqs.~\eqref{eq:dtl1} and~\eqref{eq:dtl2} are still at least reasonable. 
For~$k<\bar{k}$, the flow equation~\eqref{eq:dtl0} of the four-quark coupling~$\lambda_{\text{csc}}$ then reduces to 
\be
\partial_t \lambda_{\text{csc}} = 2\lambda_{\text{csc}} + 16\lambda_{\text{csc}}^2 b_{(2,0)}\,,
\label{eq:dtl3}
\ee
where~$b_{(2,0)}<0$ for~$\mu/k\to 0$, see App.~\ref{app:tfcts}. 

The initial condition for the flow equation~\eqref{eq:dtl3} at~$k=\bar{k}$ is given by Eq.~\eqref{eq:dtl2}. Note 
that~$b_{(2,0)}$ is associated with a purely fermionic one-loop diagram with only two internal fermion lines and four external fermion lines. 

From the flow equation~\eqref{eq:dtl3} we can now obtain an estimate for the symmetry breaking scale~$\ksb$. Indeed, this scale is defined as the 
scale at which the curvature~$\epsilon_{\mu}$ of the effective potential becomes zero, i.e., the four-quark coupling~$\lambda_{\text{csc}}=h^2/(2\epsilon_{\mu})$ 
diverges at this scale. Thus, we have~$1/\lambda_{\text{csc}}(\ksb)=0$. Next, we note that, for~$k<\bar{k}$, the flow eventually enters a regime where~$\mu/k > 1$. In this regime, the 
loop diagram~$\sim\lambda_{\text{csc}}^2$ scales as~$b_{(2,0)} \sim - c_{\psi}(\mu^2/k^2)$ with~$c_{\psi}>0$ being 
a dimensionless scheme-dependent constant.\footnote{Note that the general dependence of this four-quark interaction
on~$\mu$ is scheme-independent, at least for~$\mu/k\gg 1$, 
see also Ref.~\cite{Braun:2017srn} for a discussion.}
With this at hand, we can solve Eq.~\eqref{eq:dtl3} for the symmetry breaking scale and find $\ksb \sim \bar{k} \exp(-c/\mu^2)$, 
with~$c =  \bar{k}^2/(16 c_{\psi} \lambda_{\text{csc}}(\bar{k}))>0$ being a dimensionless constant. Plugging now Eq.~\eqref{eq:dtl2} into this expression for~$\ksb$, 
we finally arrive at the following result for the symmetry breaking scale:
\be
\ksb \sim \exp\left(- \frac{\bar{c}}{g^4\mu^2} \right)\,,
\label{eq:ksbg4}
\ee
where $\bar{c} = -\bar{k}^2/(8 c_{\psi} b_{(0,4)}^{(A)})$ is a {\em positive} constant and the strong coupling is assumed to be evaluated at the scale~$\bar{k}$.\footnote{In practice, 
the scale~$\bar{k}$ should come with an implicit dependence on the chemical potential which, however, is expected to be weak for sufficiently 
small values of the chemical potential.}  
Since the symmetry breaking scale~$\ksb$ sets the scale for low-energy observables, such as the diquark gap, we conclude that~$\Delta_{\text{gap}} \sim \ksb$. 
This assumption is indeed confirmed by our numerical results, see Fig.~\ref{fig:gap}. 

We emphasize that our result for the dependence of the symmetry breaking scale~$\ksb$ on the strong coupling differs from the 
one reported in, e.g., Refs.~\cite{Evans:1998ek,Evans:1998nf,Schafer:1998na}, see also Ref.~\cite{Son:1998uk}. In these seminal studies, 
it was found that~$\Delta_{\text{gap}} \sim \ksb\sim \exp(-\bar{c}^{\prime}/(g^2\mu^2))$, where~$\bar{c}^{\prime}$ is a positive constant. 
This $g^2$-dependence is a consequence of the assumption~$\lambda_{\text{csc}}\sim g^2$. Basically, 
the latter can be traced back to a tree-level consideration of four-quark interactions as triggered 
by a one-gluon exchange. In our present work, we have taken into account loop contributions to~$\lambda_{\text{csc}}$ 
which then alter the dependence of~$\ksb$ on the strong 
coupling as given in Eq.~\eqref{eq:ksbg4}. Starting from small chemical potentials, 
this change in the dependence of~$\ksb$ on the strong coupling potentially induces a more rapid increase of the diquark gap~$\Delta_{\text{gap}}$ when the chemical potential is increased. 
In any case, the scaling behavior~\eqref{eq:ksbg4} is only valid for sufficiently small values of the quark chemical potential, as discussed above. For very large chemical potentials, the diquark gap is eventually 
expected to increase mildly according to~$\Delta_{\text{gap}}\sim \mu \exp(-\bar{c}^{\prime\prime}/g)$ 
(where~$\bar{c}^{\prime\prime}>0$ is a constant)~\cite{Son:1998uk}, 
such that~$\Delta_{\text{gap}}/\mu$ still decreases, see Ref.~\cite{Alford:2007xm} for a detailed discussion of the diquark gap at very high densities. 

We now turn to a more quantitative comparison of our present results with already existing results for the diquark gap. Of course, 
a direct comparison is difficult as it in principle requires to consider 
the diquark gap as a function of the density. Bearing this in mind, a comparison of results for the diquark gap as a function of the quark chemical 
potential can nevertheless be valuable to gain at least a qualitative understanding of the underlying dynamics.

To be specific, let us 
compare our present results for the diquark gap obtained from the computation with ``gapped" gluons with those from one of the early 
seminal model studies in this field~\cite{Alford:1997zt} 
and our recent results~\cite{Leonhardt:2019fua}, see orange band in Fig.~\ref{fig:gap}.\footnote{Here, we restrict ourselves to the case with ``gapped" gluons 
since gluonic contributions are expected to be (partially) suppressed in the low-energy regime as a consequence of the Anderson-Higgs mechanism anyhow.} 
The width of this band represents an estimate for the theoretical uncertainty in this study.
The calculations reported in Ref.~\cite{Leonhardt:2019fua} are also based 
on an analysis of RG flows of dense QCD matter starting from the underlying quark-gluon dynamics. 
Compared with our present work, however, a Fierz-complete ansatz for 
the four-quark interaction channels has been employed in Ref.~\cite{Leonhardt:2019fua}. Whereas this aspect appears
to be less relevant when the quark chemical potential becomes large 
(since the diquark channel considered in the present work 
has been found to be most dominant in this regime~\cite{Leonhardt:2019fua,Braun:2019aow}), Fierz-incompleteness becomes more and more 
of an issue when the chemical potential is decreased, see Ref.~\cite{Braun:2019aow} for a detailed discussion. 
In fact, not only the scalar-pseudoscalar interaction channel but also vector channels become relevant 
when we approach the nucleonic low-density regime~\cite{Braun:2019aow}, see also Refs.~\cite{Song:2019qoh,Pisarski:2021aoz,Tripolt:2021jtp}. 
Since we have not included such channels in our present work, our results are expected to become less reliable when the chemical potential becomes small. 
Comparing our results for the size of the diquark gap with those from Ref.~\cite{Leonhardt:2019fua} (see left panel of Fig.~\ref{fig:gap}), 
we observe that our present results exceed those from Ref.~\cite{Leonhardt:2019fua} for~$\mu/\LQCD\lesssim  2.1$. 
Therefore, we cautiously conclude that four-quark interaction channels 
other than the diquark channel become relevant in this regime.  
Note that, towards smaller chemical potentials (associated with densities~$n/n_0 \lesssim 5$, where $n_0$ is the nuclear saturation density), 
the results from the Fierz-complete study in Ref.~\cite{Leonhardt:2019fua} are remarkably consistent with those from 
low-energy models (e.g., Ref.~\cite{Alford:1997zt}), see caption of Fig.~\ref{fig:gap} and 
also Ref.~\cite{Leonhardt:2019fua} for a discussion. 
Although the range of chemical potentials studied in our present work is beyond the range of values that can be reliably studied with 
low-energy models, we may cautiously deduce from this discussion that the inclusion of gluonic contributions leads to an 
increase of the diquark gap.

In the regime associated with diquark-channel dominance, the results for the diquark gap from the aforementioned Fierz-complete calculation (see Ref.~\cite{Leonhardt:2019fua})
and our present study are remarkably consistent. Note that, in Ref.~\cite{Leonhardt:2019fua}, the ``transition" between the high-energy degrees of freedom 
and the effective low-energy degrees of freedom has been performed at a fixed scale~$\Lambda_0$. In principle, this scale should even carry a $\mu$-dependence which 
is however at least difficult to determine {\it a priori}. In any case, 
the presence of this scale introduces a systematic uncertainty in the results, as indicated by the width of the orange band. 
We emphasize that we have removed the dependence on the scale~$\Lambda_0$ in our present work 
by implementing the dynamical hadronization technique. 
In the regime associated with a diquark-channel dominance, where a direct comparison of the two studies is most meaningful, 
we observe that the use of this technique already pays off. 
Indeed, the presence of the scale~$\Lambda_0$ in Ref.~\cite{Leonhardt:2019fua} also limits the range of 
accessible quark chemical potentials, $\mu\lesssim \Lambda_0$. 
Since the transformation of high-energy degrees of freedom into 
low-energy degrees of freedom is performed continuously in our present work, the range of chemical potentials is only constrained by the requirement that 
the chemical potential should be sufficiently smaller than the initial RG scale~$\Lambda$. 

\section{Towards constraints for low-energy models of dense QCD matter}\label{sec:LEM}
\subsection{Low-energy model couplings at high density}
Let us now turn to a discussion of the IR values of the quark-diquark coupling~$h$ and the four-diquark coupling~$\lambda_{\Delta}$
which often play an important role in the construction of low-energy models of dense QCD matter.

From the right panel of Fig.~\ref{fig:gap}, we deduce that the quark-diquark coupling and the four-diquark coupling 
are smaller in the approximation with ``gapped" gluons in the low-energy regime than 
in the approximation with ungapped gluons. However, their qualitative behavior as a function 
of the chemical potential is the same in the two approximations. 
Indeed, we observe that these two couplings decrease with increasing chemical potential in both cases. 
This simultaneous decrease is in accordance with our observation that the size of the gap appears to ``saturate" for increasing chemical potential, 
as also suggested by our analytic study of the scaling behavior of the gap~$\Delta_{\text{gap}}\sim k_{\text{SB}}$, 
see Eq.~\eqref{eq:ksbg4}. In fact, 
a decrease of the four-diquark coupling~$\lambda_{\Delta}$ 
with increasing chemical potential tends to ``pull" the position of the minimum of the effective 
action to larger values. This change of the position of the minimum 
needs to be compensated by a corresponding decrease 
of the quark-diquark coupling~$h$ such that the gap $\Delta_{\text{gap}}$ ``saturates" for 
increasing chemical potential. 
From a phenomenological standpoint, the behavior of the quark-diquark coupling and the four-diquark coupling 
suggests that interactions  between quarks and diquarks as well as among diquarks 
themselves become weaker when the density is increased, indicating that QCD is effectively described by a state of 
weakly coupled color-superconducting matter at (very) high densities. 

Of course, the actual values of the quark-diquark coupling and the four-diquark coupling depend 
on the regularization scheme as specified by the regulator function in our RG flow study. 
However, the widths of the uncertainty bands shown in Fig.~\ref{fig:gap} are essentially determined by the variation of the strong coupling 
at the initial RG scale. The uncertainty arising from a variation of the 
regulator function is found to be much smaller, see our discussion in App.~\ref{app:tfcts} for details. 

\subsection{QCD-constrained low-energy model}
From the standpoint of model building, it may be beneficial to employ the results from our RG study to constrain 
existing low-energy models of dense QCD matter. In the following, we shall demonstrate this aspect by 
considering the following quark-diquark model:
\be
S_{\text{LEM}} &=& \int\text{d}^4x\,\bigg\{ \psib_a \left({\rm i}\partial\fslash \!-\! {\rm i} \mu\gamma_0\right)\psi_a
+ \frac{1}{2}\bar{\lambda}_{\text{csc}}^{-1} \bar{\Delta}^\ast_a \bar{\Delta}_a \nn\\ 
&& \qquad\qquad 
+ \frac{\bar{\lambda}_{\Delta}}{\bar{h}^4}(\bar{\Delta}^\ast_a \bar{\Delta}_a)^2  
\!+\! \frac{1}{2}  {\rm i} (\psi^T_b\CC\gamma_5\tau_2\bar{\Delta}_a \epsilon_{abc}\psi_c)\nn\\
&& \qquad\qquad\qquad
 - \frac{1}{2} {\rm i} (\psib_b\gamma_5\tau_2 \bar{\Delta}^\ast_a \epsilon_{abc}\CC \psib^T_c)
\bigg\} \,,
\label{eq:lem}
\ee
where $a,b,c$ are color indices and we have suppressed flavor indices for readability. 
The action~$S_{\text{LEM}}$ basically represents a frequently employed 
low-energy model of dense QCD matter (for reviews, see Refs.~\cite{Rajagopal:2000wf,Alford:2001dt,Buballa:2003qv,Shovkovy:2004me,Alford:2007xm}), except for 
the fact that we also allow for a four-diquark coupling. The inclusion of the latter is inspired by our RG study which suggests that four-diquark interactions 
are generated dynamically already at high scales. Therefore, such interactions should be expected to be 
present at scales of the order of the ``hadronic" scale~$\Lambda_{\text{LEM}}\sim {\mathcal O}(1\,\text{GeV})$ 
at which low-energy models are usually defined.
\begin{figure}[t]
  \includegraphics[width=\columnwidth]{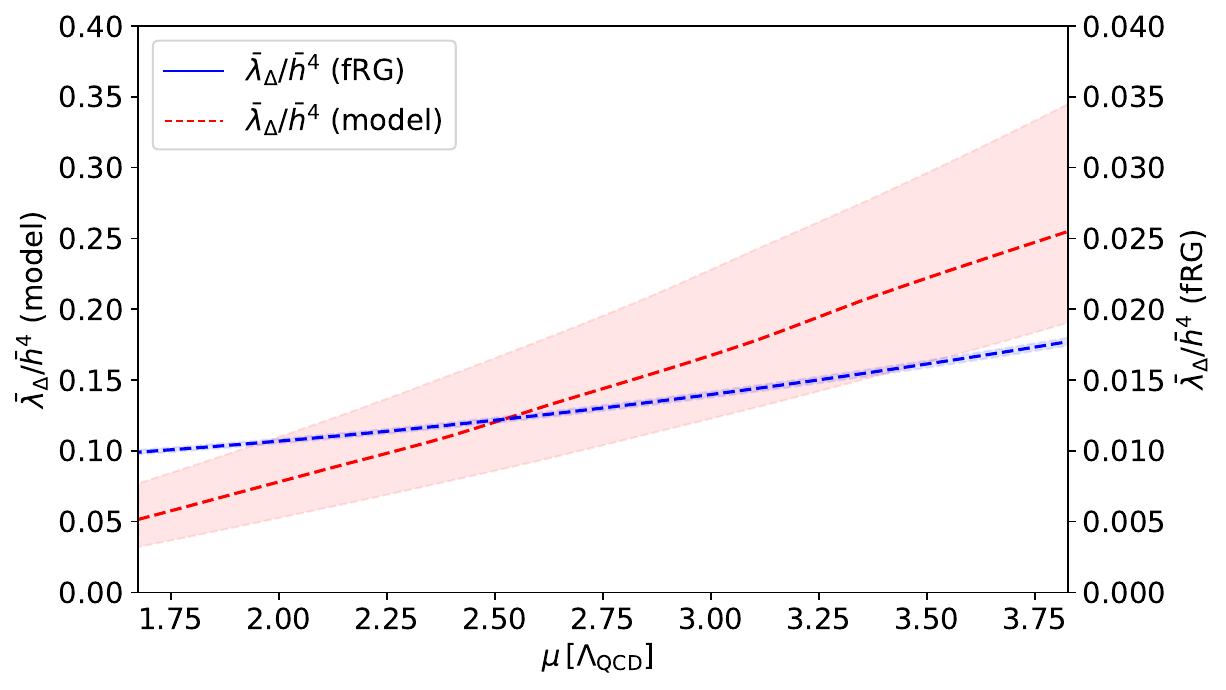}
\caption{Model parameter~$\bar{\lambda}_{\text{eff}}=\bar{\lambda}_{\Delta}/\bar{h}^4$ as a function 
of the chemical potential compared with the RG results for~$\bar{\lambda}_{\Delta}/\bar{h}^4$ 
as obtained from an evaluation of the flow at~$k=\Lambda_{\text{LEM}}=1\,\text{GeV}$. 
The shaded (blue) band associated with the RG results for~$\bar{\lambda}_{\Delta}/\bar{h}^4$ 
reflects the uncertainty arising from a variation of the strong coupling at the initial RG scale. 
In case of the model parameter, the shaded (red) band results from the uncertainty band associated with our RG estimate for the 
gap. Note that, in our model study, we adjust the 
parameter~$\bar{\lambda}_{\text{eff}}=\bar{\lambda}_{\Delta}/\bar{h}^4$ such that we recover the RG results for the gap~$\Delta_{\text{gap}}$  
as obtained in the approximation with ``gapped" gluons in the low-energy regime.}
\label{fig:mp}
\end{figure}

In the action~$S_{\text{LEM}}$ defining our model, we have introduced the fields~$\bar{\Delta}_a$ which are directly 
related to the diquark fields~$\Delta_a$ in the ansatz~\eqref{eq:gammak} for the effective action underlying 
our fRG study. We have $\bar{\Delta}_a = \bar{h}\Delta_a$. Since we shall assume that 
the quark-diquark coupling~$\bar{h}$ in our model~\eqref{eq:lem} does not depend on the RG scale~$k$, it is indeed 
convenient to rescale the original diquark fields in this way. In fact, Yukawa-type couplings such as the quark-diquark coupling 
are often treated as scale-independent quantities in low-energy model studies. In any case, the introduction of the fields~$\bar{\Delta}_a$ allows 
us to identify the coefficient of the curvature term~$\sim\bar{\Delta}^\ast_a \bar{\Delta}_a$ in Eq.~\eqref{eq:lem} with the inverse of the 
four-quark coupling~$\bar{\lambda}_{\text{csc}}$ (up to a numerical factor), 
see our discussion of the relation of the curvature and the four-quark coupling in Subsec.~\ref{subsec:dg}.
Note that, by comparing the ansatz~\eqref{eq:gammak} for the effective action underlying our fRG study 
with the action~$S_{\text{LEM}}$ of our low-energy model, we observe that
the effective action~\eqref{eq:gammak} encompasses the action~$S_{\text{LEM}}$. 

From a computation of the effective action~$\Gamma_{\text{LEM}}$ associated with the action~$S_{\text{LEM}}$, we can in principle extract 
thermodynamic quantities which are relevant for phenomenological applications. 
However, this requires to fix the parameters of the model in the first place. In the following, we shall illustrate how this can be done 
in a mean-field study of~$\Gamma_{\text{LEM}}$.
The derivation of the corresponding effective action can be found in App.~\ref{app:hdlem}.

Let us start our discussion of the determination of the model parameters 
by considering the four-quark coupling~$\bar{\lambda}_{\text{csc}}$ in Eq.~\eqref{eq:lem}.
Our analytic study of the four-quark coupling in Subsec.~\ref{subsec:dg} [in particular, see the discussion of Eqs.~\eqref{eq:dtl0}-\eqref{eq:ksbg4}] 
suggests that this coupling depends only weakly on the chemical potential, provided that we consider RG scales which are 
sufficiently large compared to the chemical potential. This is in accordance with 
our numerical results where we observe that~$\bar{\lambda}_{\text{csc}}$ evaluated at scales sufficiently greater than the chemical potential shows only a 
very mild dependence on the chemical potential. Since 
we fix the model parameters at a scale~$\Lambda_{\text{LEM}} > \mu$, we shall therefore assume that the parameter~$\bar{\lambda}_{\text{csc}}$ does not 
depend on the chemical potential. However, the value of the effective 
four-diquark coupling~$\bar{\lambda}_{\text{eff}}=\bar{\lambda}_{\Delta}/{\bar{h}^4}$ is assumed to depend 
on the chemical potential. The latter assumption is also in accordance with our RG results, see Fig.~\ref{fig:mp}. 
The actual values of the model parameters~$\bar{\lambda}_{\text{csc}}$ and~$\bar{\lambda}_{\text{eff}}$ for a given value of the 
chemical potential are finally determined by tuning them such that we recover the value of the gap~$\Delta_{\text{gap}}$ as obtained in our RG study. 
We emphasize again that we only consider~$\bar{\lambda}_{\text{eff}}$ to be $\mu$-dependent. 
The value of the four-quark coupling~$\bar{\lambda}_{\text{csc}}$ remains constant for 
all values of the chemical potential considered below. 
\begin{figure}[t]
  \includegraphics[width=\columnwidth]{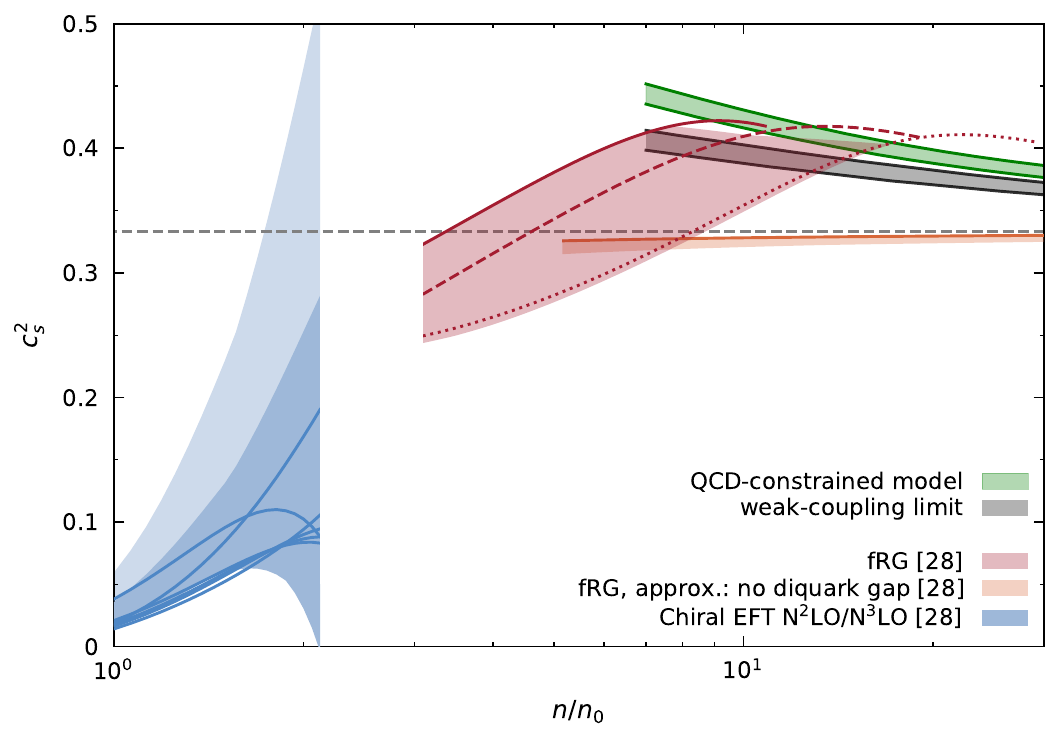}
\caption{Speed of sound squared (in units of the speed of light squared) as a function of the baryon density~$n$ (in units of the nuclear 
saturation density~$n_0$) as obtained from calculations based on chiral EFT (blue-shaded bands)~\cite{Leonhardt:2019fua}, 
an fRG study taking into account the formation of a diquark gap (red-shaded band)~\cite{Leonhardt:2019fua}, 
an fRG study based on an approximation without taking into account a diquark gap~\cite{Leonhardt:2019fua}, and
from our QCD-constrained model (green-shaded band), including the result in the weak-coupling limit (black-shaded band).
The gray dashed line is associated with the 
result for the speed of sound squared of the noninteracting quark gas.}
\label{fig:css}
\end{figure}

In the following we choose~$\Lambda_{\text{LEM}}=1\,\text{GeV} (\approx 4.8\,\Lambda_{\text{QCD}})$ 
which enables us to cover a reasonably large range of chemical potentials. For the four-quark coupling~$\bar{\lambda}_{\text{csc}}$, we
choose~$\bar{\lambda}_{\text{csc}}^{-1}\approx 0.197\,\text{GeV}^2$ (for all chemical potentials considered here).
For a given value of the chemical potential, the model parameter~$\bar{\lambda}_{\text{eff}}$ is then determined by tuning it such that the value of 
the gap in our model study agrees with the one found in our RG study. Here, we focus on the 
results for the gap as obtained in the approximation with ``gapped" gluons in the low-energy regime. However, we shall also comment on the case 
of ungapped gluons below. 

In Fig.~\ref{fig:mp}, we show the model parameter~$\bar{\lambda}_{\text{eff}}$ as a function 
of the chemical potential. There, we also present our fRG results for this quantity as obtained 
from an evaluation of the RG flow at the characteristic model scale~$k=\Lambda_{\text{LEM}}$.  
From this we deduce that the dependence of the model parameter on the chemical potential is 
compatible with our fRG results. Indeed, in both cases, we observe an increase with increasing chemical potential. 
Note that this is also the case when we evaluate the RG flow at lower scales.
Finally, we add that a larger value of the four-quark coupling~$\bar{\lambda}_{\text{csc}}$ 
requires to choose larger values of~$\bar{\lambda}_{\text{eff}}$ 
to ensure that the gap~$\Delta_{\text{gap}}$ as a function of the chemical potential remains unchanged.

\subsection{Thermodynamics}
We now use our QCD-constrained model to estimate the speed of sound of dense QCD matter. 
To this end, we first consider the pressure~$P$ as obtained from 
the effective action~$\Gamma_{\text{LEM}}$ evaluated at the ground state (gs):
\be
P= - \frac{1}{V_4}\Gamma_{\text{LEM}}\Big|_{\text{gs},\mu} + P_0\,.
\ee
Here,~$V_4$ is the spacetime volume. The determination of the (vacuum) constant~$P_0=(1/V_4)\Gamma_{\text{LEM}}|_{\text{gs},\mu=0}$ 
requires to compute the ground state in the vacuum. In QCD, the ground state 
is governed by spontaneous chiral symmetry breaking in the low-density regime.
Since we only take into account diquark-like interaction channels (which have been found to be most dominant at high 
densities~\cite{Leonhardt:2019fua,Braun:2019aow}), the low-density regime is not reliably accessible 
in our present study. However, at higher densities, derivatives of the pressure with 
respect to the chemical potential are accessible. A phenomenologically relevant quantity 
of this kind is the speed of sound~$c_s$:
\be
c_s = \frac{1}{\sqrt{\mu}}  \left(\frac{\partial P}{\partial \mu}\right)^{\frac{1}{2}} \left(  \frac{\partial^2 P}{\partial \mu\partial\mu} \right)^{-\frac{1}{2}}\,.
\label{eq:css}
\ee
By solving the baryon density~$n$, 
\be
n = \frac{1}{3}\frac{\partial P}{\partial\mu}\,,
\ee
for the chemical potential~$\mu$, we can then  
compute the speed of sound as a function of the density. 

In Fig.~\ref{fig:css}, we compare the speed of sound squared as a function of the density 
as obtained from our QCD-constrained model with results from a previous fRG study and 
calculations based on chiral EFT interactions at low densities. The green-shaded band associated 
with our model study originates from the uncertainty in the gap, see Fig.~\ref{fig:gap}. 
Starting at high densities, we find that the speed of sound increases 
with decreasing density. In particular, the speed of sound is found to be greater than the one of the noninteracting quark gas in the considered density regime. 
Note that our present estimate for the speed of sound is in reasonable agreement with the one from Ref.~\cite{Leonhardt:2019fua} for~$n/n_0\gtrsim 7$. 
This is essentially the density regime where the diquark interaction channel 
has been found to be most dominant in a Fierz-complete study~\cite{Leonhardt:2019fua,Braun:2019aow}. For lower densities, the dynamics is governed 
by chiral interaction channels and therefore this regime is not accessible in our present analysis. Still, the behavior of the 
speed of sound at high densities observed in our present study 
and the one found at low(er) densities in Ref.~\cite{Leonhardt:2019fua} (chiral EFT and fRG) 
suggests the existence of a maximum in the speed of sound for~$n/n_0 \lesssim 10$. Of course, a more accurate determination 
of the speed of sound from the full fRG flow presented in this work -- rather than from our ``QCD-constrained model" --  is in order and will be presented elsewhere. 
Based on our previous studies~\cite{Leonhardt:2019fua,Braun:2019aow}, such a calculation then also requires the inclusion of the chiral dynamics. 

We rush to add that we have also analyzed the dependence of our results 
on our choice for the model parameters. Indeed, we have 
some freedom in the model parameters since we adjust two parameters, $\bar{\lambda}_{\text{csc}}$ and~$\bar{\lambda}_{\text{eff}}$, to reproduce 
one quantity, namely the gap. Importantly, we find that the dependence on the actual choice for the parameters~$\bar{\lambda}_{\text{csc}}$ and~$\bar{\lambda}_{\text{eff}}$ 
is only mild and does not alter the qualitative behavior of the speed of sound as a function of the density, 
provided that the parameters are tuned such that the gap remains unchanged as a function of the chemical potential. 

A change of the size of the gap as a function of the chemical potential affects the speed of sound. For example, the 
model parameters can also be adjusted such that we recover the gap obtained in the fRG calculations with 
ungapped gluons in the low-energy regime, which is significantly greater than the one found in the approximation 
with ``gapped" gluons (see Fig.~\ref{fig:gap}). This results in an increase 
of the speed of sound squared of up to 70\% towards the lower end of the considered density range. However, the qualitative dependence of the 
speed of sound (squared) as a function of the density is not altered, i.e., it still increases when 
the density is decreased.

The robustness of our results for the speed of sound 
with respect to a variation of the model parameters becomes at least plausible by considering the 
weak-coupling limit of the effective action which is analytically accessible. In this limit of weak four-quark and 
four-diquark coupling, the pressure reads~\cite{Rajagopal:2000wf,Rajagopal:2000ff,Shovkovy:2002kv,Braun:2018svj}:
\be
P =  P_{\text{SB}} \left( 1 + \frac{2|\Delta_{\text{gap}}|^2}{\mu^2} + \dots\right) \,.
\label{eq:Pwclimit}
\ee
Here,~$P_{\text{SB}} = \mu^4/(2\pi^2)$ is the pressure of the 
noninteracting quark gas,\footnote{Gluons do not
contribute to~$P_{\text{SB}}$ in the zero-temperature limit.} i.e., the pressure in the so-called Stefan-Boltzmann (SB) limit.  
Interestingly, the expression~\eqref{eq:Pwclimit} does not exhibit an explicit dependence 
on the model parameters. It only depends on the chemical potential and the gap, which is a physical observable.\footnote{In principle, 
this expression is also encompassed in our present fRG study anchored in QCD since it follows from 
a consideration of the weak-coupling limit of the one-loop approximation of the effective action, see Ref.~\cite{Braun:2018svj}. 
Recall that we have~$\lambda_{\text{csc}}\sim g^4$ for the four-quark coupling, see Eq.~\eqref{eq:dtl2}. A detailed discussion of this aspect 
will be presented elsewhere. In any case, in QCD with two massless quark flavors at high density, the gap sets the scale. It is therefore 
reasonable to expect that the pressure in units of the pressure of the noninteracting quark gas can be expanded 
in powers of the dimensionless quantity~$|\Delta_{\text{gap}}|/\mu$.} 
This expression may therefore be associated 
with regimes where~$|\Delta_{\text{gap}}|/\mu$ is sufficiently small. Of course, 
the gap depends implicitly on the model parameters, such as the four-quark coupling, as also suggested by our analytic study of the scaling behavior 
of the symmetry breaking scale~$k_{\text{SB}}$ and the gap~$\Delta_{\text{gap}}\sim k_{\text{SB}}$, see Eq.~\eqref{eq:ksbg4}.  
Moreover, we observe that the leading-order correction to the Stefan-Boltzmann limit is quadratic in~$|\Delta_{\text{gap}}|/\mu$. Thus,
it increases by, e.g., a factor of four when the gap is increased by 
a factor of two for a given chemical potential.

Plugging now our fRG results for, e.g., the gap obtained 
in the approximation with ``gapped" gluons into the expression~\eqref{eq:Pwclimit} for the pressure, we can estimate the 
speed of sound with the aid of Eq.~\eqref{eq:css}. Recall that the gap in our RG study is generated from the fundamental 
quark-gluon dynamics and it therefore depends on the strong coupling~$g$. This is made explicit in Eq.~\eqref{eq:ksbg4}.   
In any case, reassuringly, we find again that the speed of sound exceeds 
the value of the noninteracting quark gas and increases when the density is decreased, see Fig.~\ref{fig:css}. 
The width of the associated black-shaded band in Fig.~\ref{fig:css} results from the width of the band of the gap shown in Fig.~\ref{fig:gap}. 
Note that we have~$0.3 \lesssim |\Delta_{\text{gap}}|/\mu \lesssim 0.6$ in the considered density range.

It is also worth adding that the observed behavior of the speed of sound as a function of the density 
has not been observed in fRG calculations which do not take into account the formation of a gap at high densities, see Ref.~\cite{Leonhardt:2019fua}
and Fig.~\ref{fig:css} for an illustration.

In accordance with Ref.~\cite{Leonhardt:2019fua}, we therefore cautiously conclude from our analysis that the
appearance of a maximum in the speed of sound
-- which exceeds the value of the noninteracting quark gas -- appears to be tightly 
connected to the formation of a diquark gap. 
Our present analysis suggests that the maximum appears in the regime~$n/n_0 \lesssim 10$ for isospin-balanced QCD matter, 
although the determination of its exact position requires additional more advanced studies, as already indicated above. With respect to astrophysical 
applications, it is still worth mentioning that the analysis of constraints 
from neutron-star masses also strongly suggests the existence of a maximum in the speed of sound for neutron-rich matter~\cite{Bedaque:2014sqa,Tews:2018kmu,Greif:2018njt,Annala:2019puf,Huth:2020ozf}.  
In any case, our present findings may already provide useful information for future studies of thermodynamic quantities at supranuclear 
densities and also for the further development of existing models of dense QCD matter. 

\section{Conclusions}
\label{sec:conc}
Starting from the fundamental quark and gluon degrees of freedom in the high-energy regime, we have studied the dynamical formation 
of diquarks in the low-energy regime at high densities, with the strong coupling at the initial RG scale as the only input parameter. 
With the present work, we have therefore laid the methodological foundation which will enable us to provide updates of our recent 
computation of the EOS of dense QCD matter~\cite{Leonhardt:2019fua}.} In particular, 
we have successfully demonstrated that the dynamical hadronization technique allows us to remove 
the dependence of an auxiliary scale~$\Lambda_0$ used in Ref.~\cite{Leonhardt:2019fua} to parametrize  
the ``transition" between the (effective) degrees of freedom at high and low energies. 
Moreover, this technique allows us to extend our 
studies to (very) high densities, even beyond the densities discussed in Ref.~\cite{Leonhardt:2019fua}. As a first application, 
we computed the diquark gap over a wide range of chemical potentials. 
We also combined these methodological advances with the implementation of a recently developed class of regulators, which 
is well suited for studies of relativistic theories in the presence of a Cooper instability~\cite{Braun:2020bhy}. 

The comparison of our present work with our previous studies~\cite{Leonhardt:2019fua,Braun:2019aow} turned out to be very beneficial, 
also for future computations of the EOS of dense matter. For example, approaching the nucleonic low-density regime from 
high densities (associated with large quark chemical potentials), this comparison indicates that the use of a Fierz-complete basis 
of (gluon-induced) four-quark interactions becomes more and more relevant. At high densities, where the diquark interaction channel 
is most dominant, a Fierz-incomplete ansatz including only the diquark channel in the matter sector 
appears to be a reasonable approximation in terms of the number of included quark interaction channels. 

Our study of the RG flow of dense QCD matter allowed us to analyze the dependence of the size of the diquark gap 
on the strong coupling and the quark chemical potential.
Moreover, we have discussed that the inclusion of gluon screening effects in our calculations may become 
particularly relevant at (very) high densities. We argued that such effects may even 
lead to a decrease of the symmetry breaking scale and the diquark gap for some intermediate range of the chemical potential, 
before they eventually increase again~\cite{Son:1998uk,Alford:2007xm}.  
In any case, it appears 
reasonable to expect that the inclusion of gluon screening effects in our calculations will render the symmetry breaking scale and the diquark gap smaller. 
However, these effects should become subleading when the chemical potential is decreased.

In addition to gluon screening effects, the inclusion of fluctuations of the diquark fields is important. 
Above the symmetry breaking scale, this is straightforward but is expected to be subleading anyhow. In fact, the corresponding contributions 
are parametrically suppressed by large screening masses of the diquarks in this regime. Below the symmetry breaking scale, however, the situation is more involved. 
Here, the inclusion of diquark fluctuations requires to deal with an Anderson-Higgs-type mechanism in future studies, which eventually leads to a suppression of gluonic contributions to the RG flow. 

Finally, we add that we have demonstrated how our present fRG study may already be used to further develop existing models
of dense QCD matter. Based on this, we have presented an analysis of a quantity which is of great interest for phenomenological applications, namely 
the speed of sound. Starting at high densities, our results indicate an increase of this quantity when the density is decreased, suggesting 
the existence of a maximum in the speed of sound of isospin-balanced QCD matter at supranuclear densities. 
This maximum would exceed the asymptotic value of the 
speed of sound associated with the limit of a noninteracting quark gas. Moreover, our study indicates that the actual height of this maximum may be sensitive to the
actual size of the gap in the fermionic excitation spectrum. 
These observations may also be interesting with respect to astrophysical applications where the equation of state of QCD matter  
enters as an input. Note that the existence of a maximum in the speed of sound of neutron-rich matter is strongly supported by the 
analysis of constraints from neutron-star masses~\cite{Bedaque:2014sqa,Tews:2018kmu,Greif:2018njt,Annala:2019puf,Huth:2020ozf}.

It is clear that our present study can and should be improved in various directions. Still, we believe that it already provides an important insight 
into the dynamics of dense QCD matter. Very importantly, our present work sets the methodological stage that allows us to connect the perturbative high-energy regime 
associated with quarks and gluons with the non-perturbative low-energy regime governed by the emergence of \mbox{(color-)}superconducting ground states at high densities. By 
successively implementing the aforementioned extensions in our present study, we expect that it will be possible to systematically improve our recent 
prediction for the EOS of nuclear matter over a wide range of densities~\cite{Leonhardt:2019fua}.

{\it Acknowledgments.--} 
The authors would like to thank A.~Gei\ss el, K.~Hebeler, M.~Leonhardt, and J.~M.~Pawlowski for useful discussions and comments on the manuscript. 
Moreover, J.B. acknowledges useful discussions with J.~Berges. 
As members of the fQCD collaboration~\cite{fQCD}, the authors also would like to thank the other members of this collaboration for discussions and 
providing data for cross-checks. 
J.B. acknowledges support by the DFG under grant BR~4005/4-1 and BR~4005/6-1 (Heisenberg program).
This work is supported by the Deutsche
Forschungsgemeinschaft (DFG, German Research Foundation) -- Projektnummer
279384907 -- SFB 1245.


%
\appendix
\section{Threshold functions and regulator}
\label{app:tfcts}
In this Appendix, we list the so-called threshold functions which appear in our RG flow equations 
and correspond to 1PI Feynman diagrams. These functions also encode
the regularization scheme dependence. The regularization scheme is determined by 
so-called regulator functions for the fermionic and bosonic fields, respectively. These functions 
are constructed such that they suitably modify the dispersion relation of the associated particles for any finite~$k$ and 
disappear in the limit~$k\to 0$. 

In our present work, we employ so-called spatial regulators and integrate out fermionic fluctuations 
around the Fermi surface which is suitable in the presence of a Cooper instability, see Ref.~\cite{Braun:2020bhy} for 
a detailed discussion. To this end, it is convenient to introduce quasi-particle dispersion relations for the fermions:
\be
\epsilon_\pm=\left(\mu \pm\left| \vec{p}^{\,}\right|\right)(1+r_\pm)\,.
\ee
These relations depend on the regularization scheme which, in our case, is specified by the following 
so-called fermionic regulator-shape functions:
\be
r_\pm:=r_{\psi}\left(x_\pm\right)\,,
\ee
where~$x_{\pm}=(\mu\pm\left| \vec{p}^{\,}\right|)^2/k^2$.
For convenience, we have used shape functions of the following form:
\be
r_\psi(x_{\pm})=-1+\frac{1}{\sqrt{1-\left(\sum\nolimits_{n=0}^{N}\frac{1}{n!}x^n_{\pm}\right)^{-1}}}\,.
\label{eq:rpsidefapp}
\ee
These functions cut off the spatial momenta exponentially in the limit~$N\to \infty$. To be specific, we have 
used~$N=4,6,8$ in our numerical calculations to analyze the regularization-scheme dependence of our results. 

For the gauge fields, we have employed 
the corresponding bosonic version of this class of regulators:
\be
r_A(x)=\frac{1}{\sum\nolimits_{n=1}^{N}\frac{1}{n!}x^{n}}\,, 
\label{eq:rAdefapp}
\ee
where $x=\vec{p}^{\,2}/k^2$. In this work, we have used~$N=4,6,8$ as for the fermionic regulator. 

With these definitions at hand, let us now define the threshold functions entering our RG flow equations:
\be
b_{(0,4)}^{(1)}(\tilde{\mu},\tilde{\chi})&=& \int_{-\infty}^{\infty} \frac{{\rm d}y_0}{2\pi} \int \frac{{\rm d}^3y}{(2\pi)^3}
\tilde{\partial}_t\bigg\{\,\tilde{G}_A^2\tilde{G}_{+}\tilde{G}_{-}\bigg(\,1 
\nn\\ 
&&\qquad 
 - \frac{29}{40}\tilde{\chi}\bigg(\frac{1 }{y_0^2 + \tilde{\epsilon}_-^2 + \tilde{\chi}} 
+ \frac{1 }{y_0^2 + \tilde{\epsilon}_+^2 +  \tilde{\chi}}\bigg)
\nn\\
&&
\!\!\!\!\!\! +\frac{3}{4}\tilde{\chi}^2\frac{1}{(y_0^2 + \tilde{\epsilon}_+^2 +  \tilde{\chi} )}\frac{1 }{(y_0^2 + \tilde{\epsilon}_-^2 +  \tilde{\chi})}\bigg)\bigg\}\,, 
\label{eq:b041}
\ee
where $y_0=p_0/k$,~$\vec{y}=\vec{p}/k$,~$\tilde{\epsilon}_{\pm}= {\epsilon}_{\pm}/k$, and~$\tilde{\chi}$ is a parameter associated with the diquark gap in our case. 
Note that~$\vec{y}^{\,2}=x$. 
Moreover, we have introduced the operator~$\tilde{\partial}_t$:
\be
\!\!\!\!\!\!\!\!\!\!\! \tilde{\partial}_t = (\partial_t r_A - \eta_A r_A)\partial_{r_A} + (\partial_t r_{+})\partial_{r_{+}} 
+ (\partial_t r_{-})\partial_{r_{-}}\,. 
\ee
The functions~$\tilde{G}_A$ and~$\tilde{G}_{\pm}$ are defined as
\be
\!\!\!\!\!\!\!\!\!\!\!\!\!\!\!\tilde{G}_A = \frac{1}{y_0^2 + \vec{y}^{\,2} (1+r_A)}
\quad\text{and}\quad
\tilde{G}_{\pm}= - \frac{1}{y_0 + {\rm i} \tilde{\epsilon}_{\pm}}\,,
\ee
respectively. In addition to the threshold function defined in Eq.~\eqref{eq:b041}, the following threshold functions appear in our RG flow equations 
for the curvature $\epsilon_{\mu}$, the diquark condensate~$\kappa$, the quark-diquark coupling~$h$, and 
the four-diquark coupling~$\lambda_{\Delta}$:  
\be
b_{(0,4)}^{(2)}(\tilde{\mu},\tilde{\chi}) 
&=&\frac{1}{2}
 \int_{-\infty}^{\infty} \frac{{\rm d}y_0}{2\pi} \int \frac{{\rm d}^3y}{(2\pi)^3}
\tilde{\partial}_t\bigg\{
\tilde{G}_A^2 \left(\frac{1}{y_0^2+ \tilde{\epsilon}_+^2 
 + \tilde{\chi}} 
 \right.
 \nn\\
&& \qquad\qquad\qquad\quad  \left.   
+\frac{1}{y_0^2+ \tilde{\epsilon}_-^2+ \tilde{\chi}} \right)
\bigg\}\,,
\ee
\vspace{-0.1mm}
\be
b_{(0,4)}^{(3)}(\tilde{\mu},\tilde{\chi}) 
&=&  \int_{-\infty}^{\infty} \frac{{\rm d}y_0}{2\pi} \int \frac{{\rm d}^3y}{(2\pi)^3}
\tilde{\partial}_t\bigg\{ \tilde{G}_A^2\bigg( \frac{1}{(y_0^2+ \tilde{\epsilon}_-^2+ \tilde{\chi})^2} 
\ \nn\\
&& 
\qquad\qquad\;
+   6\frac{1}{(y_0^2+ \tilde{\epsilon}_-^2+ \tilde{\chi})}\frac{1}{(y_0^2+ \tilde{\epsilon}_+^2+ \tilde{\chi})} 
\nn\\ 
&& \qquad\qquad\quad 
+  \frac{1}{( y_0^2+ \tilde{\epsilon}_+^2+ \tilde{\chi})^2 } \bigg) \tilde{\chi} \bigg\}\,,
\ee
\be
b_{(0,4)}^{(A)}(\tilde{\mu},\tilde{\chi}) 
&=& 
\frac{5}{36} b_{(0,4)}^{(1)}(\tilde{\mu},\tilde{\chi})  +\frac{4}{9}b_{(0,4)}^{(2)}(\tilde{\mu},\tilde{\chi}) 
 \nn \\
&& \qquad 
+  \frac{5}{48} b_{(0,4)}^{(3)}(\tilde{\mu},\tilde{\chi}) \,,
\ee
\begin{align}
b_{(1,2)}^{(A)}(\tilde{\mu},\tilde{\chi}) 
&=
 \frac{1}{2}
 \int_{-\infty}^{\infty} \frac{{\rm d}y_0}{2\pi} \int \frac{{\rm d}^3y}{(2\pi)^3}
\tilde{\partial}_t\bigg\{
\tilde{G}_A\bigg(\frac{1}{y_0^2+ \tilde{\epsilon}_+^2+ \tilde{\chi}} 
\nn \\ \nn \\
& \qquad\qquad\qquad\; 
+\frac{1}{y_0^2+ \tilde{\epsilon}_-^2+ \tilde{\chi}}     \bigg) 
\bigg\}\,,
\label{eq:b12A}
\end{align}
\begin{align}
b_{(4,0)}(\tilde{\mu},\tilde{\chi}) 
&= \frac{1}{2} \int_{-\infty}^{\infty} \frac{{\rm d}y_0}{2\pi} \int \frac{{\rm d}^3y}{(2\pi)^3}
\tilde{\partial}_t
\bigg\{ \frac{1}{(y_0^2 +\tilde{\epsilon}_-^2+ \tilde{\chi})^2} 
\nn\\ 
& \qquad\qquad\qquad
+ \frac{1}{(y_0^2 + \tilde{\epsilon}_+^2+ \tilde{\chi})^2} \bigg\}\,, 
\label{eq:b40A}
\end{align}
\be
b_{(2,0)}(\tilde{\mu},\tilde{\chi}) 
&=&
\frac{1}{2} \int_{-\infty}^{\infty} \frac{{\rm d}y_0}{2\pi} \int \frac{{\rm d}^3y}{(2\pi)^3}
\tilde{\partial}_t
\bigg\{\frac{1}{y_0^2 + \tilde{\epsilon}_-^2 +\tilde{\chi} }  
\nn\\
&& \qquad\qquad\qquad 
+\frac{1}{y_0^2 + \tilde{\epsilon}_+^2 +\tilde{\chi} }\bigg\}\,.
\ee
The threshold function appearing in the anomalous dimension of the diquark field is defined as follows:
\begin{align}
d_{(2,0)}(\tilde{\mu},\tilde{\chi}) 
&= 
\frac{1}{2}
 \int_{-\infty}^{\infty} \frac{{\rm d}y_0}{2\pi} \int \frac{{\rm d}^3y}{(2\pi)^3}
\tilde{\partial}_t \bigg\{\frac{2 y_0^2}{ (y_0^2 + \tilde{\epsilon}_\pm ^2  + \tilde{\chi})^3}
 \nn\\
& \qquad\qquad
+\frac{1}{2 \left(y_0- {\rm i}\tilde{\epsilon}_\pm \right)^2
  \left(y_0^2 + \tilde{\epsilon}_\pm ^2 + \tilde{\chi} \right)}
   \nn\\
& \quad
  - \frac{ 3 y_0 + {\rm i} \tilde{\epsilon}_\pm }{2 (y_0+ {\rm i} \tilde{\epsilon}_\pm )
  (y_0^2  + \tilde{\epsilon}_\pm ^2 + \tilde{\chi})^2} 
\bigg\}\,.
\label{eq:d20}
\end{align}
Note that we have defined the anomalous dimension of the diquark field via the 
second derivative of the corresponding loop diagram with respect to the zeroth 
component of the external four-momentum. Alternatively, we could have defined it via the 
second derivative with respect to the external spatial momentum. 
In the presence of a finite chemical potential, it is in principle necessary to take into account the wavefunction renormalization factors resulting from 
 both definitions since the chemical potential distinguishes the zeroth component of the four-momentum from the 
 spatial components. Nevertheless, for~$k\gg \mu$, the running obtained from the two definitions should eventually 
be the same. However, a residual difference in the running 
of the two wavefunction renormalizations remains even for $k\gg \mu$ when a three-dimensional 
regulator is employed, as it is the case in this work. This difference 
can be traced back to the fact that three-dimensional regulators break Lorentz invariance, see Ref.~\cite{Braun:2009si}. 
In the present study, we do not aim at resolving these issues. In fact, as mentioned above, we only
consider the wavefunction renormalization factor obtained from taking derivatives
with respect to the zeroth component of the external four-momentum since it is the one associated with the direction 
in momentum space distinguished by the 
chemical potential. In addition, from a practical point of view, it should be noted that the computation of this wavefunction renormalization 
is simplified by the fact that three-dimensional regulators do not depend on the zeroth component of the four-momentum.
\begin{figure}[t]
  \includegraphics[width=0.434\columnwidth]{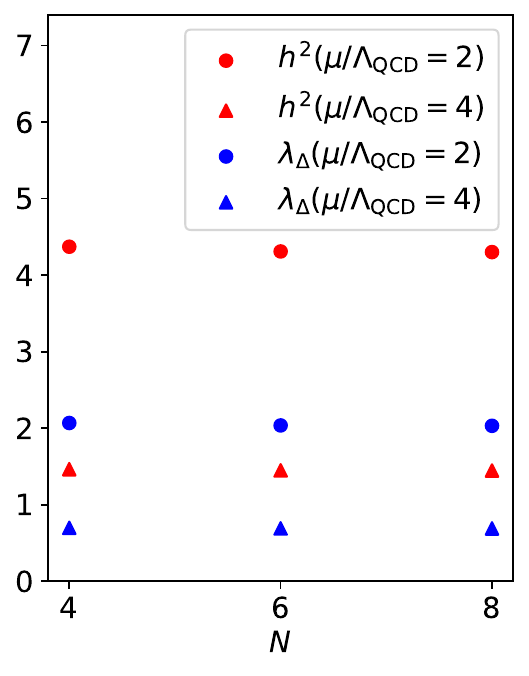}
   \includegraphics[width=0.5\columnwidth]{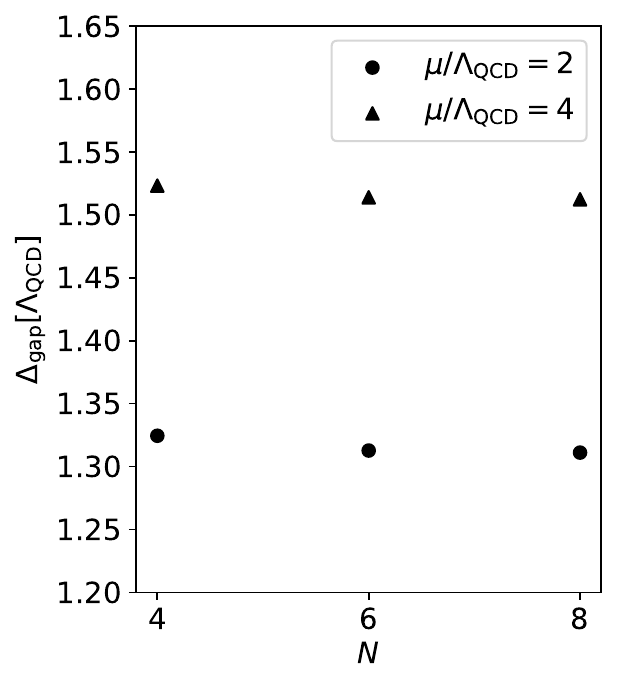}
\caption{Regulator dependence of the IR values of the 
quark-diquark coupling~$h$, the four-diquark coupling~$\lambda_{\Delta}$, 
and the gap~$\Delta_{\text{gap}}$, as specified by the parameter~$N$ for two values of the chemical potential, 
see Eqs.~\eqref{eq:rpsidefapp} and~\eqref{eq:rAdefapp}.}
\label{fig:regdep}
\end{figure}

Finally, we define the threshold function associated with the quark contribution~$\eta_{q}$ to the anomalous dimension~$Z_A$
of the gauge fields, see Eq.~\eqref{eq:etaq_gauge}. 
To this end, we first consider the RG flow equation for~$Z_A$: 
\begin{widetext}
\be
\partial_t Z_A 
&=& \frac{1}{3}\frac{\partial}{\partial Q^2_0} P_T^{\alpha\beta}(Q)
\frac{\delta_{cd}}{N_\text{c}^2-1}\frac{1}{(2\pi)^4\delta^{(4)}\left(0\right)}\frac{\delta}{\delta A_\alpha^c(Q)}\frac{\delta}{\delta A_\beta^d(-Q)}
\partial_t\Gamma_k \Bigg|_{Q_0=0,\vec{Q}=0,\Delta_a = \Delta_0\delta_{a,3}}\,.
\label{eq:ZAapp}
\ee
\end{widetext}
Here,~$\Delta_0$ is the diquark condensate,~$Q_0$ and~$\vec{Q}$ are external momenta, and~$P_T^{\alpha\beta}(Q)$ is the 
standard transversal projector. Note that the condensate~$\Delta_0$ distinguishes a direction in color space and therefore the wavefunction renormalization~$Z_A$ 
is in principle no longer uniform in color space. In our calculations, we have 
not resolved the different directions in the low-energy regime (i.e., in the 
presence of a condensate) since 
a careful analysis of this aspect requires to deal with the Anderson-Higgs mechanism. This is beyond the 
scope of this work. Below the symmetry breaking scale, our present definition of~$Z_A$ as given by Eq.~\eq{eq:ZAapp}
rather ``averages" over all directions in color space.
In any case, the threshold function associated with Eq.~\eqref{eq:etaq_gauge} is defined as
\be
d_{(2,0)}^{(A)}(\tilde{\mu},\tilde{\chi}) = -\frac{4}{\bar{g}^2}\partial_t Z_A\Big|_{\text{quark loop}} \,,
\ee
where~$\tilde{\chi}=\bar{h}^2\Delta_0^2/k^2$. It is worthwhile to add that this threshold function 
can be written in a compact form in the symmetric regime~($\tilde{\chi}=0$):
\be
\!\!\!\!\!\!\!\!\!\!\!\!\!\!\! d_{(2,0)}^{(A)}(\tilde{\mu},0) 
\!=\! \frac{32}{3}\int_{-\infty}^{\infty}\! \frac{{\rm d}y_0}{2\pi} \!\int \frac{{\rm d}^3y}{(2\pi)^3}  \tilde{\partial}_t \left\{ \!\tilde{G}_{+} \! \left(\tilde{G}_{-}\right)^3\!  \right\}.
\ee
For~$\tilde{\mu}=0$, we find~$d_{(2,0)}^{(A)}(0,0) = 2/(3\pi^2)$. By plugging this into Eq.~\eqref{eq:etaq_gauge}, we recover the one-loop result for the quark contribution to 
the running of the strong coupling in case of two massless quark flavors, as it should be. 

We close this appendix on the threshold functions and the regulator by noting 
that the uncertainty bands given in Fig.~\ref{fig:gap} include the variation 
of our results arising from a variation of the regulator as parametrized by the value of~$N$, see Eqs.~\eqref{eq:rpsidefapp} 
and~\eqref{eq:rAdefapp}. However, 
we observe that the dependence on the regulator (scheme) is much smaller than the one introduced by
the variation of the initial value of the strong coupling. In the case of ``gapped" gluons in the low-energy regime, the 
weak regulator dependence is illustrated in Fig.~\ref{fig:regdep} for the IR values of the Yukawa coupling, the four-diquark coupling, and the gap 
for~$\mu/\Lambda_{\text{QCD}}=2.0$ and~$\mu/\Lambda_{\text{QCD}}=4.0$.

\section{Dynamical Hadronization}
\label{app:dynhad}

In this work, we employ the so-called dynamical hadronization technique to study the RG flow from 
the perturbative high-energy limit down to the low-energy regime which may be conveniently described by 
effective degrees of freedom, such as pions at low density and diquarks at high density. 

Loosely speaking, we use this technique to implement continuous Hubbard-Stratonovich 
transformations in the RG flow which map quark selfinteraction channels onto diquark interaction channels as well as quark-diquark channels. 
For example, this allows us to conveniently resolve momentum dependences of, e.g., four-quark interactions and to compute 
the order-parameter potential. In general, this technique is even more powerful as it relies on the idea of 
introducing scale-dependent fields~\cite{Gies:2001nw,Gies:2002hq,Pawlowski:2005xe,Gies:2006wv,Floerchinger:2009uf,Braun:2014ata,Fu:2019hdw}. 

In the present work, we introduce $k$-dependent diquark fields, $\Delta_{a}\rightarrow\Delta_{a,k}$. The original Wetterich equation~\cite{Wetterich:1992yh} is 
then modified as follows:
\begin{widetext}
\be
\partial_t\Gamma_k = \partial_t\Gamma_k\Big|_{\Delta_{a,k},\Delta^{\ast}_{a,k}}+\int_{p^{\prime}}\,\left\{\frac{\delta\Gamma_k}{\delta\Delta_{a,k}(p^{\prime})}\partial_t\Delta_{a,k}(p^{\prime}) 
+\frac{\delta\Gamma_k}{\delta\Delta_{a,k}^{\ast}(p^{\prime})}\partial_t\Delta^{\ast}_{a,k}(p^{\prime})\right\}\,.
\label{eq:dynhadWetterich}
\ee
\end{widetext}
The first term on the right-hand side of this equation is nothing but the original Wetterich equation evaluated on the scale-dependent diquark fields. 
For the parametrization of the scale dependence 
of these fields, we make the following ansatz:
\begin{widetext}
\be
\partial_t\Delta_{a,k}(p^{\prime})&=& - \int \frac{{\rm d}^4 p}{(2\pi)^4}\int \frac{{\rm d}^4 q}{(2\pi)^4} \frac{\rm i}{2}(\psib_b(p)\gamma_5\tau_2 \epsilon_{abc} {\mathcal C}\psib^T_c(q)) \delfourn{p^{\prime}\!-\! p\! -\! q} \partial_t\rho_k(p^{\prime}) 
 +\Delta_{a,k}(p^{\prime})\partial_t\beta_k(p^{\prime})\,,
 \label{eq:dqtf1app}
 \\ 
\partial_t\Delta_{a,k}^{\ast}(p^{\prime})&=&  \int \frac{{\rm d}^4 p}{(2\pi)^4}\int \frac{{\rm d}^4 q}{(2\pi)^4} \frac{\rm i}{2} (\psi^T_b(p){\mathcal C}\gamma_5\tau_2\epsilon_{abc}\psi_c(q)) \delfourn{p^{\prime}\!-\!p\!-\!q} \partial_t\rho_k(p^{\prime}) 
+\Delta_{a,k}^{\ast}(p^{\prime})\partial_t\beta_k(p^{\prime})\,,
 \label{eq:dqtf2app}
\ee
\end{widetext}
where~$p=\{p_0,\vec{p}^{\,}\}$ and correspondingly for~$p^{\prime}$ and~$q$.  
The functions~$\beta_k$ and $\rho_k$ are at our disposal. In the following, we shall determine them such that 
quark selfinteractions associated with the diquark channel are mapped onto diquark interaction channels and quark-diquark interaction channels, 
as it is usually done by a Hubbard-Stratonovich transformation of 
the quark bilinears appearing on the right-hand sides of Eqs.~\eqref{eq:dqtf1app} and~\eqref{eq:dqtf2app}. 
However, as we allow for a scale dependence in our parametrization, we perform such a transformation continuously as a function of the RG scale~$k$. This is 
important as four-quark interactions usually removed by such a Hubbard-Stratonovich transformation at a given scale may be regenerated in the RG flow because of processes associated with, e.g., two-gluon exchange diagrams. 

For the chiral regime at low densities, the determination of the functions~$\beta_k$ and $\rho_k$ has been discussed in Refs.~\cite{Gies:2001nw,Gies:2002hq,Gies:2006wv,Braun:2014ata}. Similar to these studies, 
we determine these functions by requiring that
\begin{itemize}
\item[(i)] the RG flow equation of the four-quark coupling
vanishes identically 
on all scales~$k$, 
\item[(ii)] the quark-diquark coupling~$\bar{h}$ is momentum-independent,
\item[(iii)] $\partial_t Z_{\Delta}(p_0=0, |\vec{p}^{\,}|=k) = -\eta_{\Delta} Z_{\Delta}$.
\end{itemize}
Our initial condition for the four-quark coupling, \mbox{$\bar{\lambda}_{\text{csc}}\to 0$} for~$k\to\Lambda$ (see Subsec.~\ref{subsec:sf}), 
together with the requirement~(i) ensures that a four-quark interaction channel (as associated with the 
diquark channel) is not generated in the RG flow. The contributions to this four-quark interaction generated in the RG flow are mapped 
onto the scalar sector, in the spirit of a Hubbard-Stratonovich transformation. 
The requirement~(ii) ensures that the diquark gap generated in the low-energy regime is also momentum-independent. 
Finally, our third requirement renders our approximation of a momentum-independent~$Z_{\Delta}$-factor self-consistent. 

By plugging our ansatz~\eqref{eq:gammak} for the effective action together with our ans\"atze~\eqref{eq:dqtf1app} and~\eqref{eq:dqtf2app} for 
the diquark fields into the flow equation~\eqref{eq:dynhadWetterich} and then 
applying the aforementioned three requirements, we obtain the following equations for~$\beta_k$ and~$\rho_k$ in the symmetric regime:
\be
\partial_t\beta_k(p) &=& -\frac{Z_\Delta p^2+ 4{\rm i} Z_\Delta \mu p_0 + \Delta\bar{m}^2 }{\bar{h}^2}\partial_t\bar{\lambda}_\text{csc}(p) \nn\\
&& \; +\frac{1}{Z_\Delta k^2\bar{h}^2}\Big[\left(Z_\Delta k^2 \!+\! \Delta\bar{m}^2  \right)^2 \partial_t\bar{\lambda}_\text{csc}(0,|\vec{p}^{\,}|\!=\! k) \nn \\ 
&& \; - (\Delta\bar{m}^2)^2 ( \partial_t \bar{\lambda}_\text{csc}(0,|\vec{p}^{\,}|\!=\!k) \!-\! \partial_t \Delta\bar{\lambda}_{\text{csc}} )
\Big]\,,
\ee
where~$\Delta \bar{m}^2 = \bar{m}^2 - 4 Z_{\Delta}\mu^2$, 
and
\be
\partial_t\rho_k(p)&=& \frac{1}{\bar{h}}\partial_t\bar{\lambda}_\text{csc}(p)\big|_{\Delta_k,\Delta^{\ast}_k}\,.
\ee
The quantity~$\Delta\bar{\lambda}_{\text{csc}}$ is the difference 
of~$\bar{\lambda}_\text{csc}(0,|\vec{p}^{\,}|\!=\! k)$ and~$\bar{\lambda}_\text{csc}(0,|\vec{p}^{\,}|\!=\! 0)$. 
In our study, we set this quantity to zero. This approximation has been discussed in Refs.~\cite{Gies:2001nw,Gies:2002hq}. 
There, it has been found that quantities such as the symmetry breaking scale and the condensate  
are only weakly affected by this simplification. 
Essentially, it only affects the position of (pseudo-)fixed points 
of the couplings in the symmetric regime but not their existence.

In the low-energy regime, associated with a nontrivial ground state, the equation for~$\rho_k$ remains unchanged. However, 
the equation for~$\beta_k$ changes and reads 
\be
\partial_t\beta_k(p)&=& -\frac{Z_\Delta p^2+ 4{\rm i} Z_\Delta \mu p_0}{\bar{h}^2}\partial_t\bar{\lambda}_\text{csc}(p) \nn\\
&& \qquad +\frac{Z_\Delta k^2}{\bar{h}^2}\partial_t\bar{\lambda}_\text{csc}(k)\,.
\ee
With these equations for~$\beta_k$ and~$\rho_k$ at hand, the flow equations for the couplings 
presented in Subsec.~\ref{subsec:rgfloweqs} can be computed.

\section{High-density low-energy model}\label{app:hdlem}

In this appendix, we derive the effective action~$\Gamma_{\text{LEM}}$ for the model defined in 
Eq.~\eqref{eq:lem} in Sec.~\ref{sec:LEM}. More specifically, we shall 
compute the effective action~$\Gamma_{\text{LEM}}$ in a one-loop approximation where we only take into account the purely
fermionic loop and set the wavefunction renormalizations associated with the diquark fields to zero. 
The wavefunction renormalizations of the quarks are assumed to be constant.
Note that our derivation follows closely the one of a related model 
in Ref.~\cite{Braun:2018svj}. 

The starting point is the classical action~$S_{\text{LEM}}$ of our model:
\be
S_{\text{LEM}} &=& \int\text{d}^4x\,\bigg\{ \psib_a \left({\rm i}\partial\fslash \!-\! {\rm i} \mu\gamma_0\right)\psi_a
+ \frac{1}{2}\bar{\lambda}_{\text{csc}}^{-1} \bar{\Delta}^\ast_a \bar{\Delta}_a \nn\\ 
&& \qquad\qquad 
+ \frac{\bar{\lambda}_{\Delta}}{\bar{h}^4}(\bar{\Delta}^\ast_a \bar{\Delta}_a)^2  
\!+\! \frac{1}{2}  {\rm i} (\psi^T_b\CC\gamma_5\tau_2\bar{\Delta}_a \epsilon_{abc}\psi_c)\nn\\
&& \qquad\qquad\qquad
 - \frac{1}{2} {\rm i} (\psib_b\gamma_5\tau_2 \bar{\Delta}^\ast_a \epsilon_{abc}\CC \psib^T_c)
\bigg\} \,.
\label{eq:lemapp}
\ee
Here, $a,b,c$ are color indices. The flavor indices are suppressed for readability. 

Using, e.g., the Wetterich equation~\cite{Wetterich:1992yh} and expanding the diquark fields about a
homogeneous background, we obtain the following result for~$\Gamma_{\text{LEM}}$:
\be
\frac{1}{V_4}\Gamma_{\text{LEM}} =\frac{1}{V_4}\Gamma_{\text{LEM},\Lambda} \! -\! \frac{\mu^4}{6\pi^2}\!
-\!  8 l_0(\Lambda,|\bar{\Delta}|^2)\,,
\label{eq:effdqmapp}
\ee
where~$V_4$ is the spacetime volume and~$|\bar{\Delta}|^2=\bar{\Delta}_a^{\ast} \bar{\Delta}_a$ (summation over~$a$ is 
assumed). The contribution $\sim\mu^4$ in this expression for~$\Gamma_{\text{LEM}}$ 
originates from quarks which do not couple to the diquark fields and
therefore only appear as a ``noninteracting contribution". The quark loop integral 
is parametrized by the function~$l_k$:
\begin{align}
   l_k(\Lambda,|\bar{\Delta}|^2) 
    &=\frac{1}{2} \int \frac{{\rm d}^3p}{(2\pi)^3} \theta(\Lambda^2 \!-\! 
    \vec{p}^{\,2})\theta(\vec{p}^{\,2}\!-\! k^2) \times \nn\\
 &  \qquad\qquad \times \Big\{ \sqrt{ ( |\vec{p}^{\,}| +\mu)^2 + |\bar{\Delta}|^2}  \nn\\
  &  \qquad\qquad\qquad + \sqrt{ ( |\vec{p}^{\,}| -\mu)^2 + |\bar{\Delta}|^2}\Big\}
    \,.
\label{eq:M0}
\end{align}
Here, we have employed a sharp cutoff/regulator as often used in model studies. 
The expression for this loop diagram for general 
three-dimensional regulators can be found in Ref.~\cite{Braun:2018svj}. Within the present approximation, a 
different choice for the regulator would only change the values of the model parameters used in a 
concrete calculation (i.e.,~$\bar{\lambda}_{\text{csc}}$ and~$\bar{\lambda}_{\text{eff}}=\bar{\lambda}_{\Delta}/\bar{h}^4$) and the numerical 
prefactors associated with the counter terms in~$\Gamma_{\text{LEM},\Lambda}$, as these are scheme-dependent 
quantities.

The quantity~$\Gamma_{\text{LEM},\Lambda}$ in Eq.~\eqref{eq:effdqmapp} includes two classes of terms. First, 
it contains terms which determine the form of the effective action~$\Gamma_{\text{LEM},\Lambda_{\text{LEM}}}\simeq S_{\text{LEM}}$ 
at a given scale~$\Lambda_{\text{LEM}}<\Lambda$. Second,~$\Gamma_{\text{LEM},\Lambda}$ includes counter 
terms which ensure that~$\Gamma_{\text{LEM}}$ in Eq.~\eqref{eq:effdqmapp}
is an RG-consistent effective action, i.e., $\Lambda \partial_{\Lambda}\Gamma_{\text{LEM}}=0$ for~$\Lambda\to\infty$. 
To be specific, we have~\cite{Braun:2018svj}:
\be
\frac{1}{V_4}\Gamma_{\text{LEM},\Lambda}&=&  \frac{1}{2}\bar{\lambda}_{\text{csc}}^{-1}  |\bar{\Delta}|^2 + 
 \frac{\bar{\lambda}_{\Delta}}{\bar{h}^4}  |\bar{\Delta}|^4
\nn\\
 &&\; +8 l_{\Lambda_{\text{LEM}}}(\Lambda,|\bar{\Delta}|^2)\Big|_{\mu=0}\nn\\
 &&\;\;
 + 4\mu^2\Big(\partial_{\mu}^2 l_{\Lambda_{\text{LEM}}}(\Lambda,|\bar{\Delta}|^2)\Big|_{\mu=0}\Big)\,.
\ee
For~$\Lambda=\Lambda_{\text{LEM}}$, we find~$l_{\Lambda_{\text{LEM}}}(\Lambda,|\bar{\Delta}|^2)=0$ and we are 
left with~$\Gamma_{\text{LEM},\Lambda_{\text{LEM}}}=  V_4((1/2)\bar{\lambda}_{\text{csc}}^{-1}  |\bar{\Delta}|^2 + 
 (\bar{\lambda}_{\Delta}/\bar{h}^4)  |\bar{\Delta}|^4)$. In any case, inserting~$\Gamma_{\text{LEM},\Lambda}$ into  
 Eq.~\eqref{eq:effdqmapp}, we find
 \be
  \Lambda \partial_{\Lambda} \Gamma_{\text{LEM}} \!=\! 
- 2V_4|\bar{\Delta}|^2\mu^2 \left(\frac{\mu}{\pi\Lambda}\right)^2
\!+\! {\mathcal O}(1/\Lambda^{4})\,. 
\nn
 \ee
Thus, our low-energy model described by the effective action~$\Gamma_{\text{LEM}}$ is 
RG-consistent in a strict sense in the limit $\Lambda\to\infty$. In our numerical computations of thermodynamic 
observables discussed in Sec.~\ref{sec:LEM}, we have always ensured RG consistency by choosing sufficiently 
large values for~$\Lambda$. For a detailed discussion of this aspect, we refer the reader to Ref.~\cite{Braun:2018svj}.

\bibliography{qcd}

\end{document}